\documentclass{article}
\pdfoutput=1
\usepackage{amsmath,amssymb,epsfig,color}
\numberwithin{equation}{section}
\usepackage{hyperref}

\DeclareMathAlphabet{\mathpzc}{OT1}{pzc}{m}{it}

\title{Double-logarithms in ${\cal N}$$=$ 8 supergravity: \\
impact parameter description \\\& mapping to 1-rooted ribbon graphs} 
\author{Agust{\' \i}n Sabio Vera\\ 
\\
 Instituto de F{\' \i}sica Te{\' o}rica UAM/CSIC, Nicol{\'a}s Cabrera 15\\ 
\& U. Aut{\' o}noma de Madrid, E-28049 Madrid, Spain} 

\begin{document} 

\maketitle 

The set of double-logarithmic (DL) contributions $(\alpha \, t \ln^2 {s})^n$ to the 4-graviton amplitude in ${\cal N}$$=$ 8 supergravity (SUGRA), with $\alpha$ being the gravitational coupling and $(s,t)$ the Mandelstam invariants, is studied in impact parameter ($\rho$) representation. This sector of the amplitude shows interesting properties which shed light on the nature of quantum corrections in gravity. Besides having a convergent behaviour as $s$ increases, which is not present in ${\cal N}$$<$ 4 SUGRA theories, there exists a critical line $\rho_c(s)$ above which the Born amplitude prevails. The short distance region $\rho < \rho_c(s)$ is  dominated by the DL terms. As a consequence, when studied in terms of an eikonal approach in the forward limit, the scattering angle linked to the bending of the semiclassical trajectory of the graviton shows a transition from attractive gravity at large distances to a region at small $\rho$ characterized by a repulsive DL contribution to the gravitational potential due to the gravitino content of the theory. In the complex angular momentum plane, this DL high energy asymptotics is driven by the rightmost pole singularity of a parabolic cylinder function. The resummation of DL quantum corrections in ${\cal N}$$=$ 8 SUGRA can be understood in terms of the counting of 1-rooted maps on orientable surfaces. 

\section{Introduction}

The progress in the understanding of scattering amplitudes in gauge, gravity and string theories in the last two decades has been enormous. This activity has been driven, from a phenomenological perspective, by the need to understand the physics at colliders (mainly at the Large Hadron Collider) with high precision. From a formal viewpoint, the anti-de Sitter/conformal field theory postulated duality~\cite{Maldacena:1997re,Gubser:1998bc,Witten:1998qj} has led to an intense study of  the 
${\cal N}$= 4 supersymmetric Yang-Mills model. In this theory it has been possible to evaluate quantum corrections to scattering amplitudes at large  orders by making use of conformal symmetry and integrability~\cite{Bern:2010tq}. This has allowed steady progress in the investigation of different SUGRA amplitudes which can be calculated applying a double copy prescription to the Yang-Mills results~\cite{Bern:2002kj}. This program has been very useful in the quest to find out if supersymmetric versions of gravity can be renormalisable to all orders~\cite{Bern:1998ug,Bern:2009kd}.

As a complement to the exact calculation of amplitudes, it is also possible to extract important information about them by taking judicious kinematical limits. This allows for the construction of effective field theories whose validity can be tested against experimental, mainly collider, data. To follow this path it is needed to calculate the exact coefficients for certain classes of logarithmic terms in the amplitudes whose numerical values dominate the observables under study. Quite often these coefficients can be evaluated to all orders in the coupling of the theory and a resummation can be performed. 

This program can be applied, in particular, to the so-called multi-Regge kinematics (MRK) limit of gravitational theories. Scattering amplitudes in this context, where the scattering energy is much larger than any other Mandelstam invariant, can be written in a factorised fashion which is related to the exchange of reggeized gravitons~\cite{Grisaru:1975tb,Grisaru:1981ra}, blended with eikonal and double-logarithmic terms~\cite{Lipatov:1982vv,Lipatov:1982it,Lipatov:1991nf}. All these pieces can be handled efficiently by means of a high energy effective action proposed by Lipatov~\cite{Lipatov:2011ab}. In connection with the double copy structure~\cite{Bern:2008qj}, the graviton emission vertex in MRK is directly related to the  corresponding~\cite{BFKL1,BFKL2,BFKL3} gluon emission vertex~\cite{Lipatov:1982vv,Lipatov:1982it,Lipatov:1991nf,SabioVera:2011wy,Vera:2012ds,Johansson:2013nsa,Vera:2014tda}. 

This framework was used by Bartels, Lipatov and the author in Ref.~\cite{Bartels:2012ra} to calculate the DL contributions to the 4-graviton amplitude, $(\alpha \, t \ln^2 {s})^n$ ($\alpha$ is the gravitational coupling,  $(s,t)$ are the usual  Mandelstam invariants), in different SUGRA theories and Einstein-Hilbert gravity. The DL results in~\cite{Bartels:2012ra} are in agreement with all the known results  for ${\cal N}$-SUGRA theories in four dimensions up to two loops~\cite{BoucherVeronneau:2011qv} and offer precise predictions for these terms to all orders. Very recently, Henn and Mistlberger~\cite{Henn:2019rgj} found agreement with the third order DL prediction in~\cite{Bartels:2012ra} for ${\cal N}$$=$8 SUGRA. 

The all-orders resummation of the DL sector in the 4-graviton amplitude  in~\cite{Bartels:2012ra} revealed that ${\cal N}$= 4 SUGRA is a critical theory where the DLs are simply not present. For ${\cal N}$$<$4 SUGRAs the DL sector of the amplitudes exhibits a very fast asymptotic growth with $s$ while for ${\cal N}$$>$4 SUGRAs they converge to zero rather rapidly.  The asymptotic behaviour of the DL family of terms points towards a better ultraviolet behaviour of gravity as the number of gravitinos in the theory increases. It would be very interesting to find an argument based on symmetries to explain the behaviour of the different resummations in~\cite{Bartels:2012ra}.  For this task it is worth noting that the DL terms are only sensitive to the graviton and gravitino content of the particular theory.

It is possible to get a more precise physical picture of the DL sector.   For this it is convenient to transform the amplitude and the resummed DL series into an impact parameter representation. This is the task of the present work where the focus will be on ${\cal N}$$=$8 SUGRA, which turns out to be the simplest example to perform the calculations. 

Sec. 2 presents a brief overview of the DL resummation in momentum space in terms of a differential equation for the partial wave. The coefficients of the perturbative recursive solution can be written using results from Combinatorics Mathematics~\cite{MartinKearney:2011} on exactly solvable self-convolutive recurrences. In Sec. 3 the singularity structure  in the complex angular momentum plane of the partial wave is discussed. It is argued that the high energy asymptotic behaviour of the scattering amplitude is governed by the poles generated by the two rightmost zeros of a parabolic cylinder function. Sec. 4 is devoted to the representation of the DL resummation in impact parameter ($\rho$) space by means of an eikonal phase. This phase presents a sharp transition at $\rho = \rho_c (s) \simeq 
\sqrt{\alpha} \ln{s}$ from a Born dominated behaviour at large $\rho$ to a DL driven region when $\rho<\rho_c(s)$ and admits perturbative expansions in 
${\rho_c(s) / \rho}$ and ${\rho / \rho_c(s)}$, respectively. The associated semiclassical graviton deflection angle is Newtonian at large distances. For impact parameters below $\rho_c(s)$ it quickly approaches zero to then change sign, indicating a repulsive contribution from the DLs to the gravitational potential at very small impact parameters. In Sec. 5 an introduction to the theory of ribbon graphs is presented. A one-to-one mapping between the partial wave for the DL contributions to the 4-graviton scattering amplitude and the generating function for the number of 1-rooted ribbon graphs is found. Finally, some Conclusions are drawn.

\section{Resummation of double-logarithmic terms}

The 4-graviton scattering amplitude with helicities (++;++) in ${\cal N}$$=$ 8 SUGRA can be investigated factorising out the Born contribution in the form 
\begin{eqnarray}
{\cal A}_{4} = {\cal A}_4^{\rm Born} {\cal M}_{4}, 
\, \, \, {\cal A}_4^{\rm Born} = \kappa^2 \frac{s^3}{t u},
\, \, \, {\cal M}_{4} = 1 + \sum_{n=1}^\infty {\cal M}_{4}^{(n)}.
\label{eqnnotation}
\end{eqnarray}
$L$ is the loop order and $\kappa^2 \equiv 8 \pi G \equiv 8 \pi^2 \alpha$, with $G$ being the Newton's constant. $s=(p_1+p_2)^2$, $t=(p_1-p_3)^2$ and $u=(p_1-p_4)^2$ are the Mandelstam variables.  

Making use of an infrared (IR) regulator, $\lambda$, the exact 1-loop amplitude reads
\begin{eqnarray}
{\cal M}^{(1)}_{4}  &=& \alpha \, t \ln{\left(\frac{-s}{-t}\right)}\ln{\left(\frac{-u}{-t}\right)} \nonumber\\
&+& \alpha \, \frac{t}{2} \ln{\left(\frac{-t}{\lambda^2}\right)}
\left[\ln{\left(\frac{-s}{-t}\right)}+\ln{\left(\frac{-u}{-t}\right)}\right] \nonumber\\
&-&\alpha \frac{(s-u)}{2} \ln{\left(\frac{-t}{\lambda^2}\right)}\ln{\left(\frac{-s}{-u}\right)}.
\end{eqnarray}
In the high energy limit $s \simeq -u \gg -t$ the third term dominates the amplitude since it is proportional to $\alpha \, s$ while the first two terms carry a $\alpha \, t$ factor.  For this amplitude it is well known that the IR divergent contributions exponentiate to all orders in $\alpha$~\cite{Weinberg:1965nx}. The first term is interesting because, although suppressed by a factor of $\alpha \, t$, it also carries a strong double logarithmic (DL) dependence with the center-of-mass energy, $s$, {\it i.e.}
\begin{eqnarray}
{\cal M}^{(1)}_{4,{\rm DL}}  &\simeq& \alpha \, t \ln^2{\left(\frac{s}{-t}\right)}.
\end{eqnarray}
 Furthermore, it is $\lambda$ independent and its functional structure to all orders lies away from the theorems for exponentiation of IR divergent terms. It is then important to find the structure of these DL terms to all orders, {\it i.e.}
 \begin{eqnarray}
{\cal M}_{4,{\rm DL}} &=& 1+ \sum_{n=1}^\infty {\cal M}^{(n)}_{4,{\rm DL}}  
~=~ 1+ \sum_{n=1}^\infty {\cal C}_n \, (\alpha \, t)^n \ln^{2n}{\left(\frac{s}{-t}\right)}.
\end{eqnarray}
In Ref.~\cite{Bartels:2012ra} the coefficients ${\cal C}_n$ were investigated. The steps for the derivation are briefly explained in the following and a novel explicit representation for them is also discussed. 

The amplitude with DL accuracy can be written in terms of a Mellin transform, 
\begin{eqnarray}
{\cal M}_{4,{\rm DL}}  (s,t) &=& \int _{\delta-i\infty}^{\delta +i\infty}\frac{d\,\omega }{2\pi i}\,
\left(\frac{s}{-t}\right)^\omega \frac{ f _\omega}{\omega} \,,\,\,\delta>0\, ,
\label{M4Mellin}
\end{eqnarray}
where the contour of integration lies to the right of all the singularities of the 
$t$-channel partial wave $f_\omega$ on the complex angular momentum plane. This function admits the perturbative expansion
\begin{eqnarray}
f_\omega &=& \sum _{n=0}^\infty {\cal C}_n\,\left(\frac{\alpha t}{\omega^2}\right)^n\,.
\label{firstexp}
\end{eqnarray}

Similarly to the case of quantum electrodynamics (QED) and quantum chromodynamics (QCD)~\cite{Kirschner:1982qf,Kirschner:1982xw,Kirschner:1983di},  the DL terms can be obtained from the following evolution equation for $f_\omega$ which takes into account the contributions stemming from virtual gluons and gravitinos with transverse momentum bigger than $\sqrt{-t}$, 
\begin{eqnarray}
f_\omega &=& 1- \alpha\,  t \left(\frac{d}{d\,\omega}\,\frac{f_\omega}{\omega}
- \frac{f_\omega^2}{\omega ^2}\right).
\label{fomegqeq}
\end{eqnarray}
In a graphical representation, this equation has the form 
\begin{eqnarray}
 \hspace{-1.6cm}\parbox{15mm}{\includegraphics[width=2.2cm,angle=0]{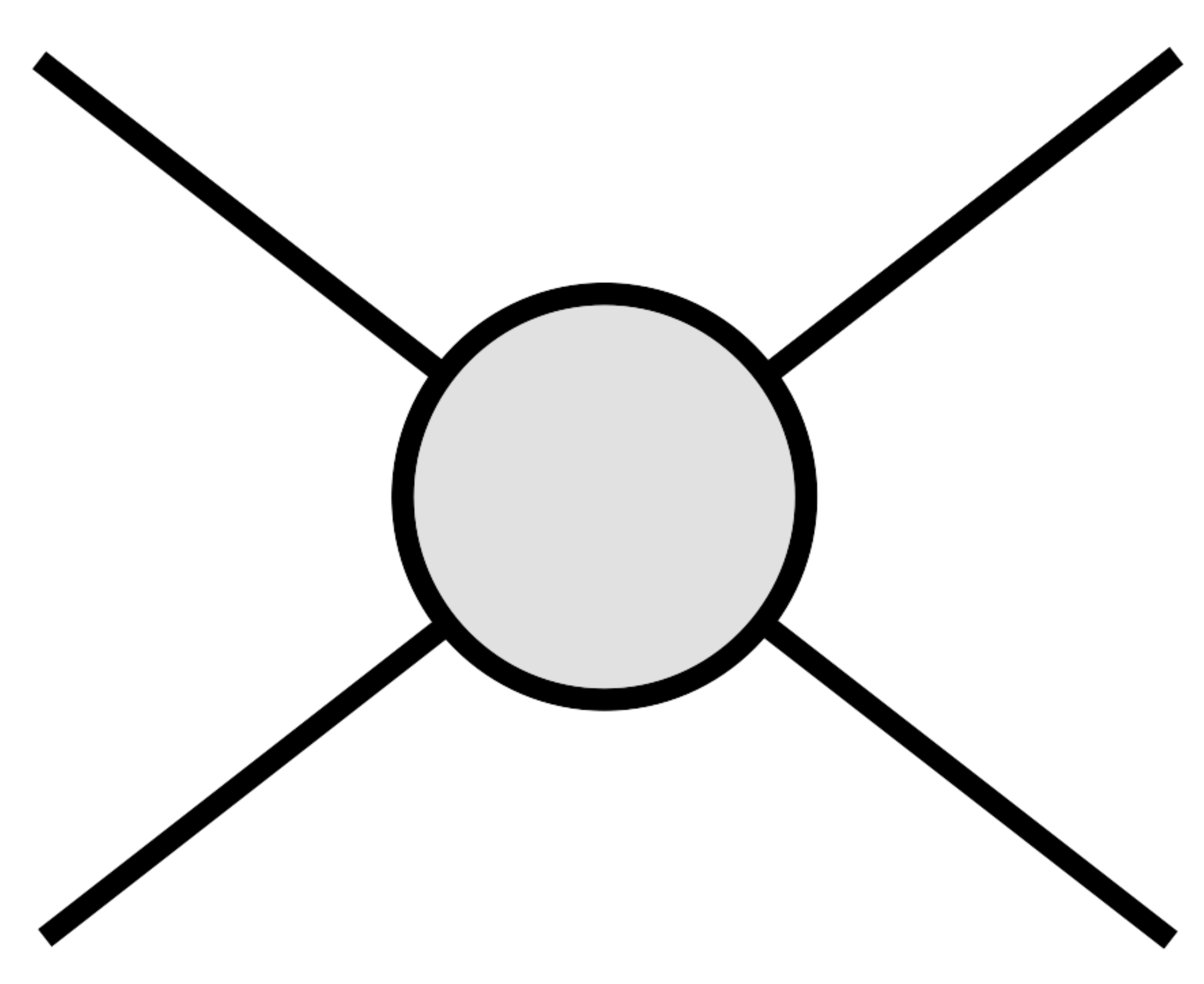}} \hspace{.4cm}
 &=& \hspace{-.3cm} \parbox{15mm}{\includegraphics[width=2.2cm,angle=0]{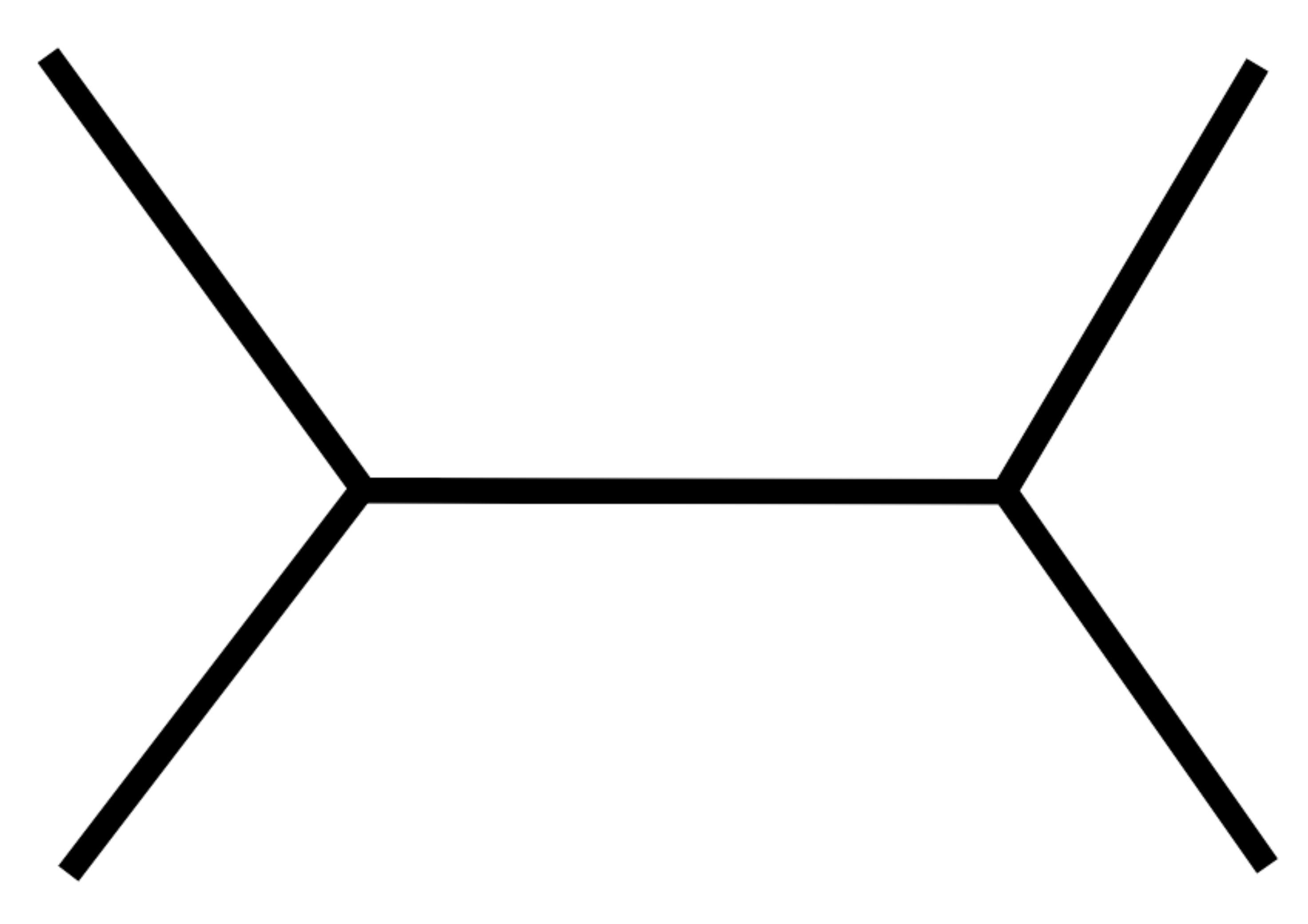}} 
 \hspace{0.6cm}
 + 2 \hspace{.cm} \parbox{15mm}{\includegraphics[width=2.2cm,angle=0]{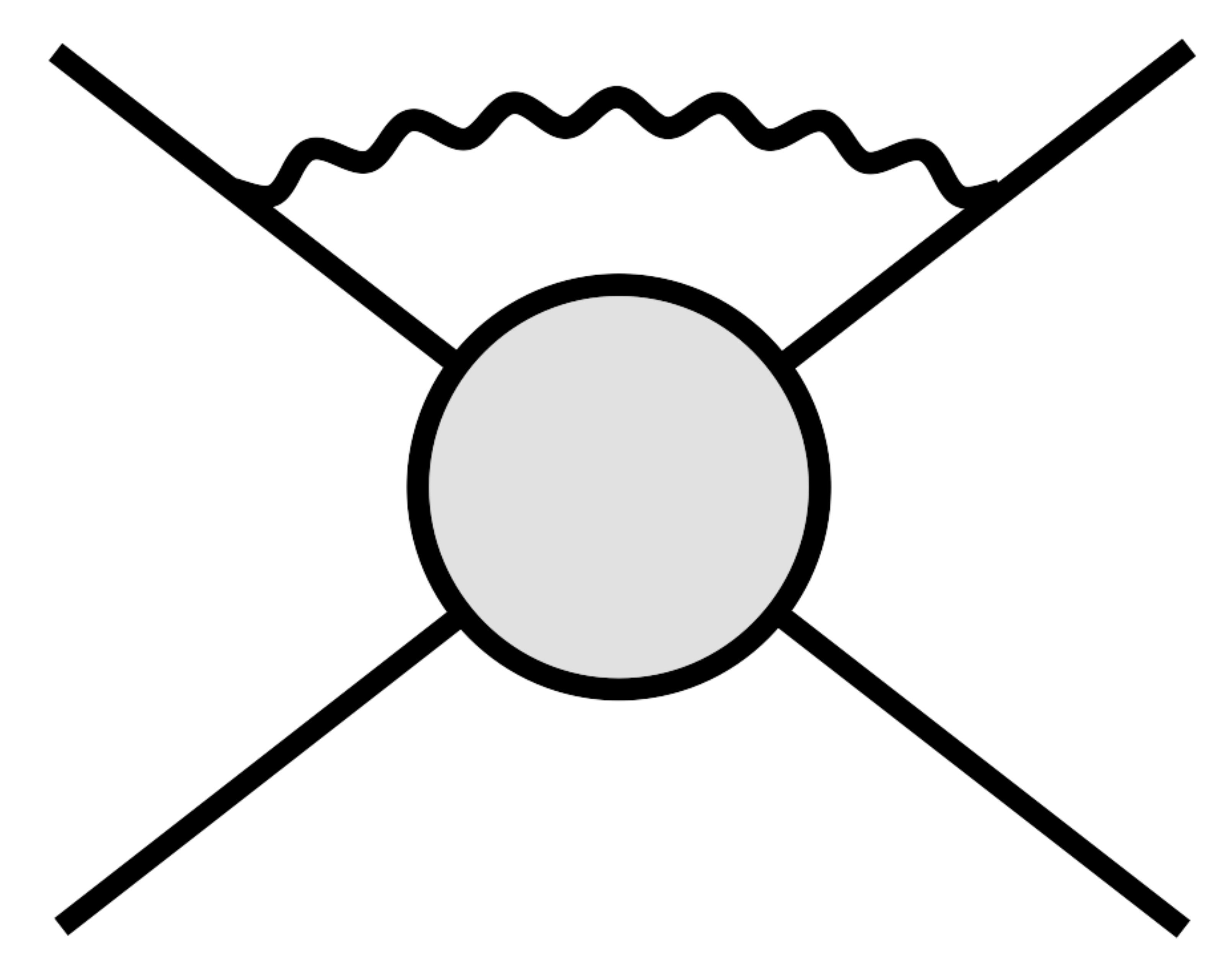}}
 \hspace{0.5cm}
 + 2 \hspace{.cm} \parbox{15mm}{\includegraphics[width=2.2cm,angle=0]{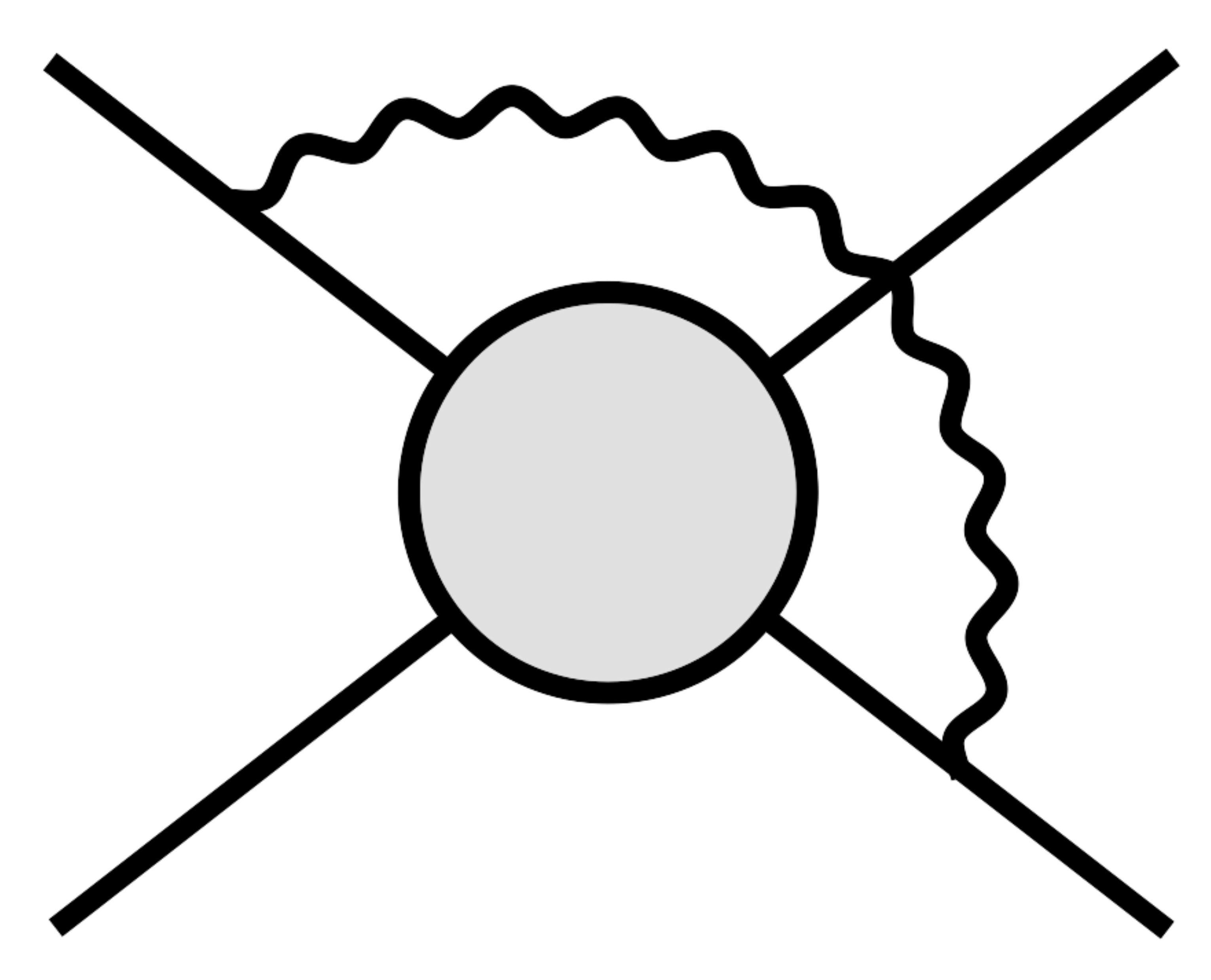}}
 \hspace{0.5cm}
  + \hspace{-.1cm} \parbox{15mm}{\includegraphics[width=2.6cm,angle=0]{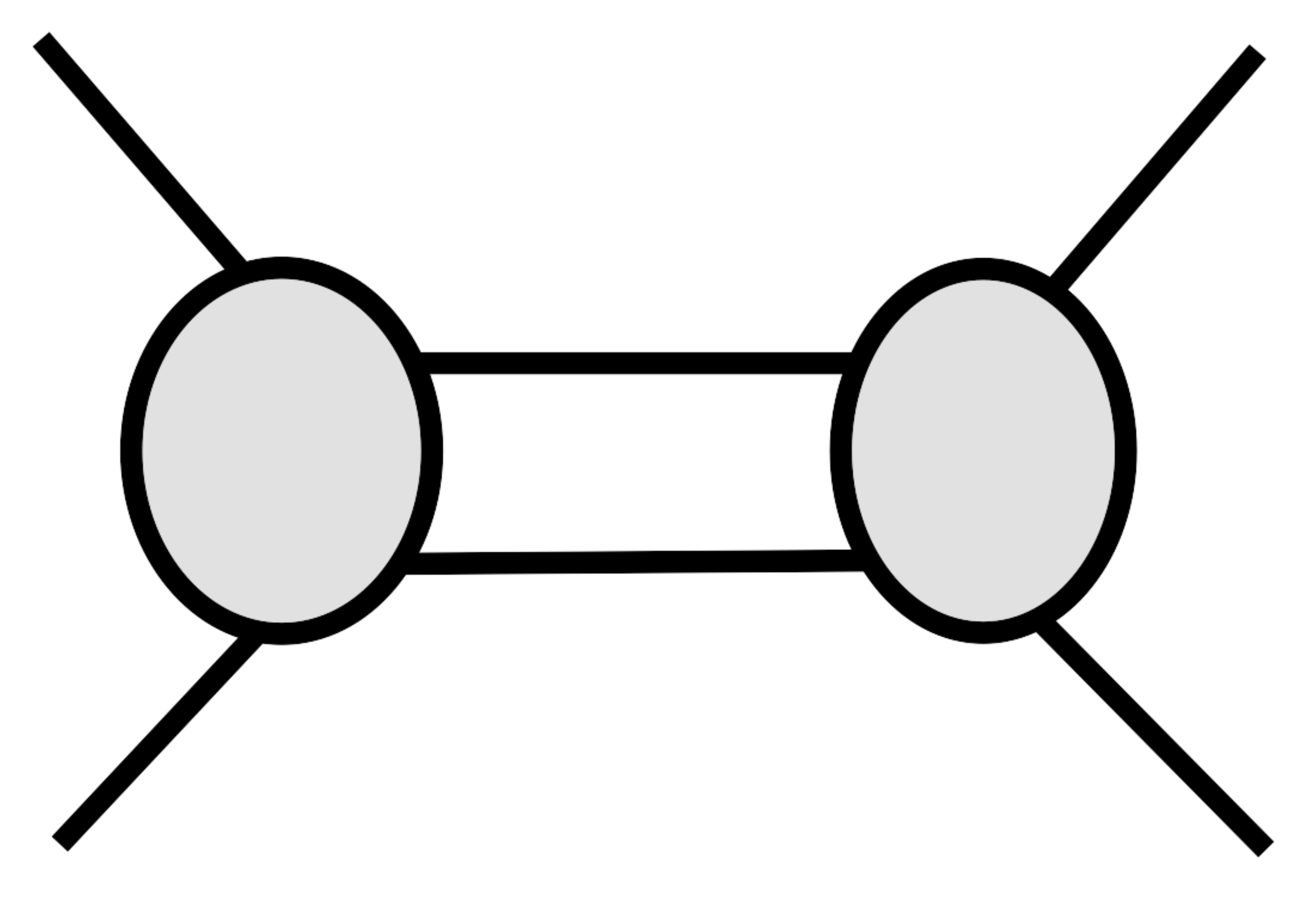}}\hspace{1cm}.\hspace{-.8cm}
\end{eqnarray}
The derivative term in Eq.~(\ref{fomegqeq}) corresponds to virtual graviton contributions with the smallest transverse momentum. The quadratic piece introduces the softest gravitons or gravitinos exchanged in the $t$-channel. At one loop the double logarithmic term arises as a contribution to the graviton Regge trajectory which in the Regge limit contains an ultraviolet divergence which can be regularized using $\sqrt{s}$ as a cut-off. The same argument applies at higher orders where, together with the ladder contributions, Sudakov-like soft graviton insertions in the external legs contribute to the evolution equation for the amplitude. 

The solution to Eq.~(\ref{fomegqeq}) can be obtained by a perturbative iteration and has the form
\begin{eqnarray}
f_\omega &=& 1 + \frac{2\,  \alpha \, t}{w^2} + \frac{10 \, \alpha^2 \, t^2}{w^4} 
+ \frac{74 \, \alpha^3 \, t^3}{ w^6}
+\frac{706 \, \alpha^4 \, t^4    }{w^8} + \frac{8162 \, \alpha^5 \, t^5 }{ w^{10}}
   + \dots
  \label{fullpertexp}
   \end{eqnarray}
The numerical coefficients ${\cal C}_n$ in this expression are present in other physical and mathematical problems. They appeared as early as 1976 in~\cite{Ihrig:1976ih} where different theorems for the  enumeration of diagrams associated to many body theory were investigated. 
 In 1978 Cvitanovi{\' c} et al. made use of field theoretical functional methods to evaluate sums of combinatoric weights of Feynman diagrams~\cite{Cvitanovic:1978wc}. In 
 QED without using Furry's theorem (which does not allow diagrams with electron loops attached to an odd number of photons) they encountered these coefficients when evaluating higher order corrections to the electron propagator. More recently, in~\cite{Goodvin:2006Go}, the Green's function for the Holstein polaron was evaluated by summing all the self-energy diagrams whose number at a given order corresponds to ${\cal C}_n$. This result is relevant to understand the coupling of electrons to lattice vibrations (phonons) in terms of a single dressed particle (polaron).  
 
Of more direct relevance for the work here presented are the results of Martin and Kearney in~\cite{MartinKearney:2011} where the class of exactly solvable self-convolutive recurrence relations of the form
\begin{eqnarray}
u_n &=& (\alpha_1 n + \alpha_2) u_{n-1} + \alpha_3 \sum_{j=1}^{n-1} u_j u_{n-j}, \,\, \, u_1~=~ 1,
\label{generalueqn}
\end{eqnarray}
were studied. It was shown that it is possible to write the solution using an  integral representation as the Mellin transfom
\begin{eqnarray}
u_n &=& \int_0^\infty x^{n-1} \mu (x) dx.
\end{eqnarray}
Introducing the expansion~(\ref{firstexp}) in Eq.~(\ref{fomegqeq}) is equivalent to Eq.~(\ref{generalueqn}) with   $(\alpha_1,\alpha_2,\alpha_3) = (2,-3,1)$ and  $u_{n+1} = {\cal C}_n$. This allows to write the partial wave in the closed form
\begin{eqnarray}
f_\omega &=&  {2 \over \sqrt{\pi}}  \sum_{n=0}^\infty 
\left(\frac{2 \alpha \, t}{w^{2}} \right)^n 
\int_0^\infty \frac{x^{n-\frac{1}{2}} e^{x}}{G^2 (x)+\pi}dx
\end{eqnarray}  
where
\begin{eqnarray}
G (x) &=& \sum_{r=0}^\infty \frac{x^{r+\frac{1}{2}}}{\left(r+\frac{1}{2}\right)r!}
~=~ \sqrt {\pi} \, {\rm erfi}\left (\sqrt {x} \right).
\end{eqnarray} 
The function ${\rm erfi} (z)$ is the imaginary error function ${\rm erf} (i z)/i$. This implies that the DL sector of the scattering amplitude is
   \begin{eqnarray}
{\cal M}_{4,{\rm DL}}  (s,t)
&=& {2 \over \pi^{3 \over 2} }  \sum_{n=0}^\infty 
\frac{\left(2 \alpha \, t \right)^n }{(2n)!} \ln^{2n} \left(\frac{s}{-t}\right) 
\int_0^\infty \frac{x^{n-\frac{1}{2}} e^{x}}{1+{\rm erfi}^2 \left (\sqrt {x} \right)} dx 
\nonumber\\
&=& 1 + 2 \left(\frac{\alpha \,  t}{2}\right) \ln^2 \left(\frac{s}{-t}\right) 
+\frac{5}{3}  \left(\frac{\alpha \,  t}{2}\right)^2 \ln^4 \left(\frac{s}{-t}\right) 
 \nonumber\\
   &+&\frac{37}{45} \left(\frac{\alpha \,  t}{2}\right)^3  \ln^6 \left(\frac{s}{-t}\right) 
   +\frac{353}{1260}  \left(\frac{\alpha \,  t}{2}\right)^4 \ln^8 \left(\frac{s}{-t}\right) \nonumber\\
   &+& \frac{583}{8100} 
    \left(\frac{\alpha \,  t}{2}\right)^5 \ln^{10} \left(\frac{s}{-t}\right)  + \dots
    \label{M4DLscamp}
\end{eqnarray}
The first three terms have been known for quite some time from calculations of the full amplitude. The higher order predictions serve as a non-trivial cross-check of the exact calculations at all orders in the gravitational coupling. The term $\sim {\cal O} (t^3 \ln^6 s)$ has recently been confirmed by Henn and Mistlberger in~\cite{Henn:2019rgj}.  The $n \to \infty$ behaviour of the coefficients ${\cal C}_n$ will be analysed in detail in the last section of this work.

 A plot of the dependence of the DL scattering amplitude of Eq.~(\ref{M4DLscamp}) with $s$ is shown in Fig.~\ref{PertResumPlotb1}.  Fifty terms have been used in the perturbative expansion and $\alpha \, t$ has been set to -1. Since the remaining of this work is devoted to the study of the DL high energy asymptotics, a simplification has been used by scaling the variable $s$ in the argument of the logarithms with a general fixed scale $\mu^2$ which is set to be 1 GeV$^2$. The replacement $\ln{s/|t|} \to \ln{s/\mu^2} \to \ln{s}$ is implemented and, for simplicity, these logarithms are denoted as $\ln{s}$ throughout the text and figures to follow.  In Fig.~\ref{PertResumPlotb1} it is worth noting the good convergence of the amplitude as $s \to \infty$. As already discussed in~\cite{Bartels:2012ra}, this convergent DL asymptotics is present in all ${\cal N}$$>$ 4 SUGRA theories but not when ${\cal N}$$<$ 4.

\section{Singularities in complex angular momentum plane}

It is possible to go beyond the perturbative analysis of the scattering amplitude by finding the solution to Eq.~(\ref{fomegqeq}) and extracting the complete singularity structure of the partial wave in the complex total angular momentum in the $t$-channel plane. The high energy asymptotics should then be dominated by those singularities situated closer to the origin of this $\omega$-plane. 

The form of the solution to Eq.~(\ref{fomegqeq})  which is compatible with the perturbative expansion in Eq~(\ref{fullpertexp}) is
\begin{eqnarray}
\frac{f_\omega}{\omega } &=&  \frac{\omega
   }{\alpha  t} + \frac{  1}{ \sqrt{-\alpha t }  \, 
   e^{\frac{-\omega ^2}{4 \alpha  t}}
     \, D_{-1} \left(\frac{\omega }{   \sqrt{-\alpha t}  }\right)}
     ~=~  
    - \frac{d}{d \omega} \ln{\left( e^{\frac{-\omega ^2}{4 \alpha  t}}D_{-1}\left(\frac{\omega}{\sqrt{- \alpha t}}\right)\right)},
    \label{NPsolfw}
\end{eqnarray}
where the parabolic cylinder function $D_{-1}$ can be written using the integral representation
\begin{eqnarray}
D_{-1} (z) = e^{-z^2 \over 4} \int_0^\infty e^{\frac{-t^2}{2}- z t} dt,\, \, \,\frac{d}{dz} D_{-1} (z) = \frac{z}{2} D_{-1} (z) - e^{-z^4 \over 4}, 
\, \, \, D_{-1}(0) = \sqrt{\frac{\pi}{2}}. 
\end{eqnarray}

To show the agreement with the perturbative series in Eq~(\ref{fullpertexp}) it is convenient to use in Eq.~(\ref{NPsolfw}) the asymptotic expansion at large $z$,
\begin{eqnarray}
D_{-1} (z) &=& \frac{e^{-z^2 \over 4} }{z}
\int_0^\infty e^{{-t^2 \over 2 z^2}- t} dt
~ \simeq ~    \frac{e^{-z^2 \over 4} }{z} \sum_{n=0}^\infty \frac{(-1)^n (2n)!}{2^n z^{2n} n!},
\label{ParabCylDexpansion}
\end{eqnarray}
while comparing both sides of
\begin{eqnarray}
\frac{   1}{\frac{f_\omega}{\omega } - \frac{\omega
   }{\alpha  t}} &=&   \sqrt{-\alpha t } \,  e^{\frac{-\omega ^2}{4 \alpha  t}}
     \, D_{-1} \left(\frac{\omega }{   \sqrt{-\alpha t}  }\right). 
\end{eqnarray}
Making use of the expansion in Eq.~(\ref{fullpertexp}) and setting 
$z= - i \omega / \sqrt{\alpha t}$ in Eq.~(\ref{ParabCylDexpansion}) both expressions lead to 
\begin{eqnarray}
&\simeq& -  2 \omega \sum_{n=0}^\infty \frac{  (2n)!}{  n!} \left({\alpha t \over 2 \omega^2}\right)^{n+1}.
\end{eqnarray}

\begin{figure}[h]
\begin{center}
\includegraphics[width=11cm]{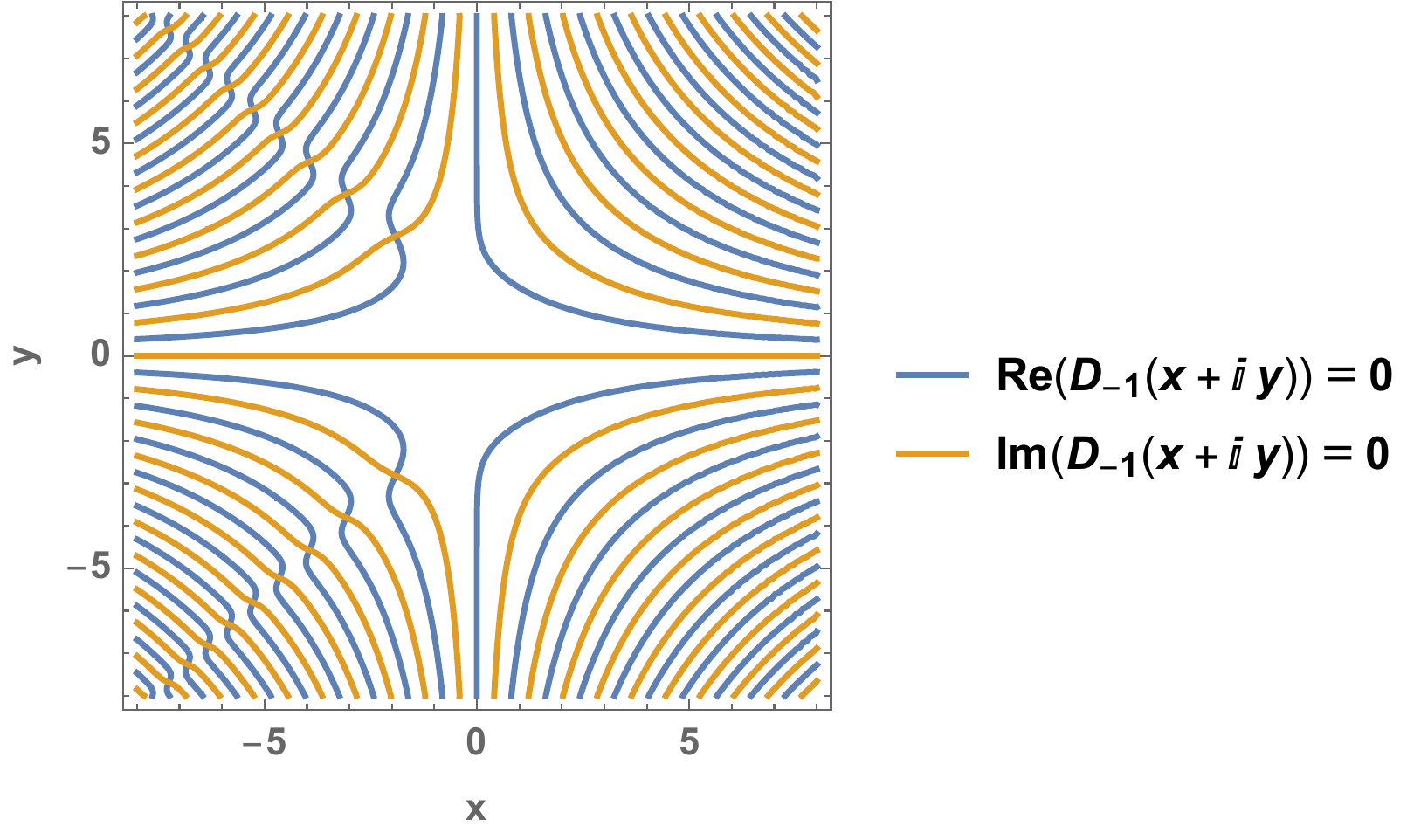}
\end{center}
\vspace{-.5cm}
\caption{Lines on the complex plane $z=x+iy$ where the real and imaginary parts of $D_{-1} (z)$  are zero.}
\label{ParabCylPlot}
\end{figure}

When performing the Mellin transform 
   \begin{eqnarray}
{\cal M}_{4,{\rm DL}}  (s,t) &=& \int_{\delta-i \infty}^{\delta + i \infty} \frac{d \omega}{ 2 \pi i} 
s^\omega \frac{f_\omega}{\omega} \nonumber\\
&=& - \int_{\delta-i \infty}^{\delta + i \infty} \frac{d \omega}{ 2 \pi i} 
s^\omega  \frac{d}{d \omega} \ln{\left( e^{\frac{-\omega ^2}{4 \alpha  t}}D_{-1}\left(\frac{\omega}{\sqrt{- \alpha t}}\right)\right)}
\label{omegaintegAmpl}
\end{eqnarray}
 to obtain the scattering amplitude, the partial wave $f_\omega$ is integrated along a vertical contour $\delta + i \nu$ with $\delta \gtrsim 0$ and $\nu \in (-\infty,\infty)$. Since  
$f_\omega/\omega$ corresponds to the logarithmic derivative of a parabolic cylinder function, it is useful to note that it develops an infinite set of simple poles at the left hand side of the complex plane which correspond to zeroes of $D_{-1} (z)$. To illustrate this point, and setting $z=x+i y$, the lines where the real and imaginary parts of $D_{-1} (z)$ are zero are shown in Fig.~\ref{ParabCylPlot}. The poles are located at the points where those lines intersect. 
\begin{figure}
\begin{center}
\includegraphics[width=6cm]{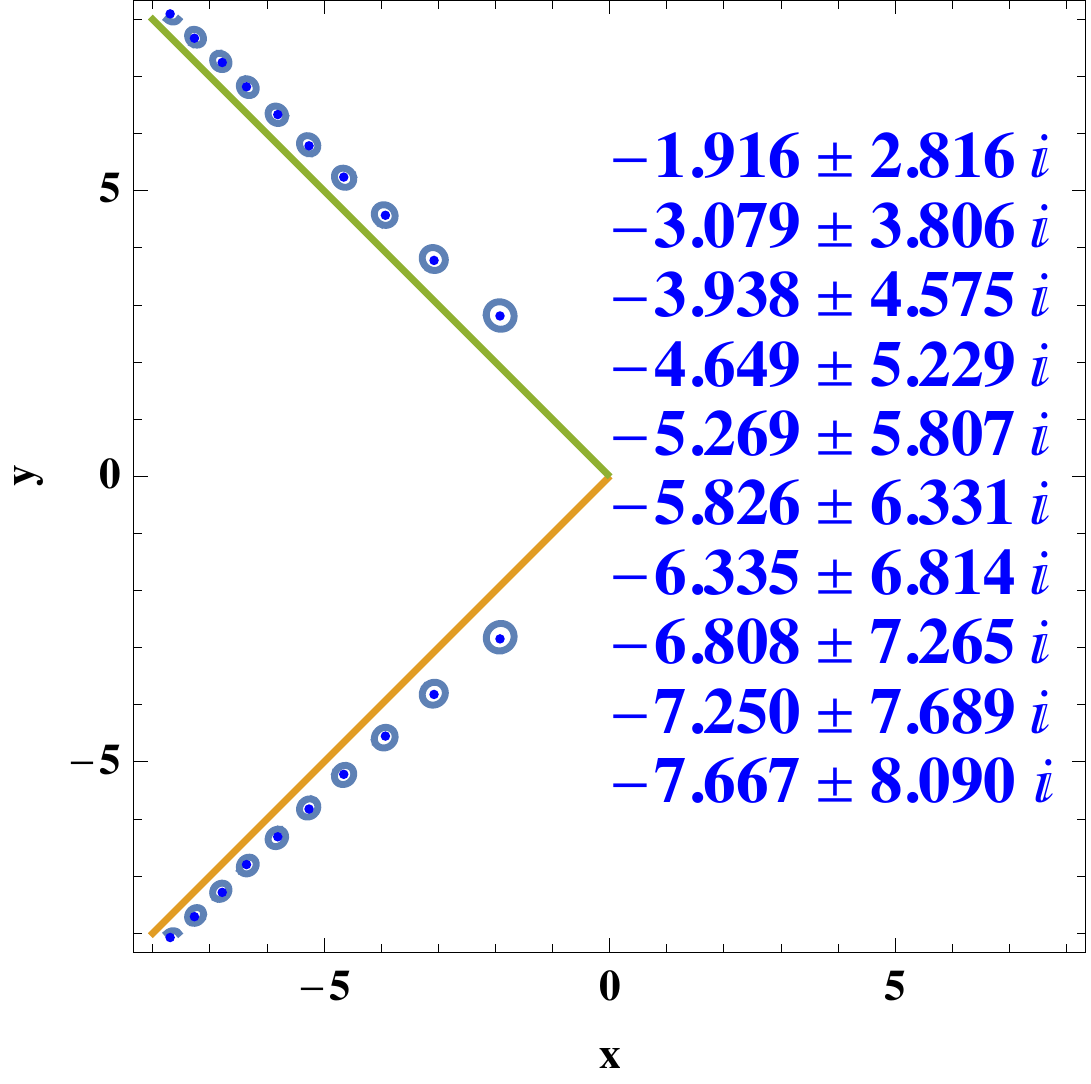}
\end{center}
\vspace{-.5cm}
\caption{The 20 zeroes on the complex plane $z=x+iy$ for $D_{-1} (z)$ with the largest real part.}
\label{ParabCylZeroesList}
\end{figure}

There are two infinite sets of complex conjugated values situated asymptotically close to the lines $\arg z=\pm \frac{3}{4}\,\pi$. The 20  zeroes with the largest real part, together with their numerical values, are shown in Fig.~\ref{ParabCylZeroesList}. As it can be understood from the Mellin transform definition in Eq.~(\ref{M4Mellin}), the DL $s \to \infty$ limit is governed by the two poles with the largest real part, $-1.916 \pm 2.816$. Deforming the contour of integration to cross them and calculating their residues generates the following asymptotic prediction for the amplitude,
\begin{eqnarray}
\lim_{s \to \infty} {\cal M}_{4,{\rm DL}}  (s,t) &=&  s^{(-1.916 + 2.816 \, i) \sqrt{- \alpha \, t}} + {\rm c.c.}\nonumber\\
&=& 2 \, s^{-1.916 \, \sqrt{- \alpha \, t}}\cos \left(2.816 \,\sqrt{- \alpha \, t}
\ln s\right).
\label{AsympAmplitude}
\end{eqnarray}

\begin{figure}
\begin{center}
\includegraphics[width=9cm]{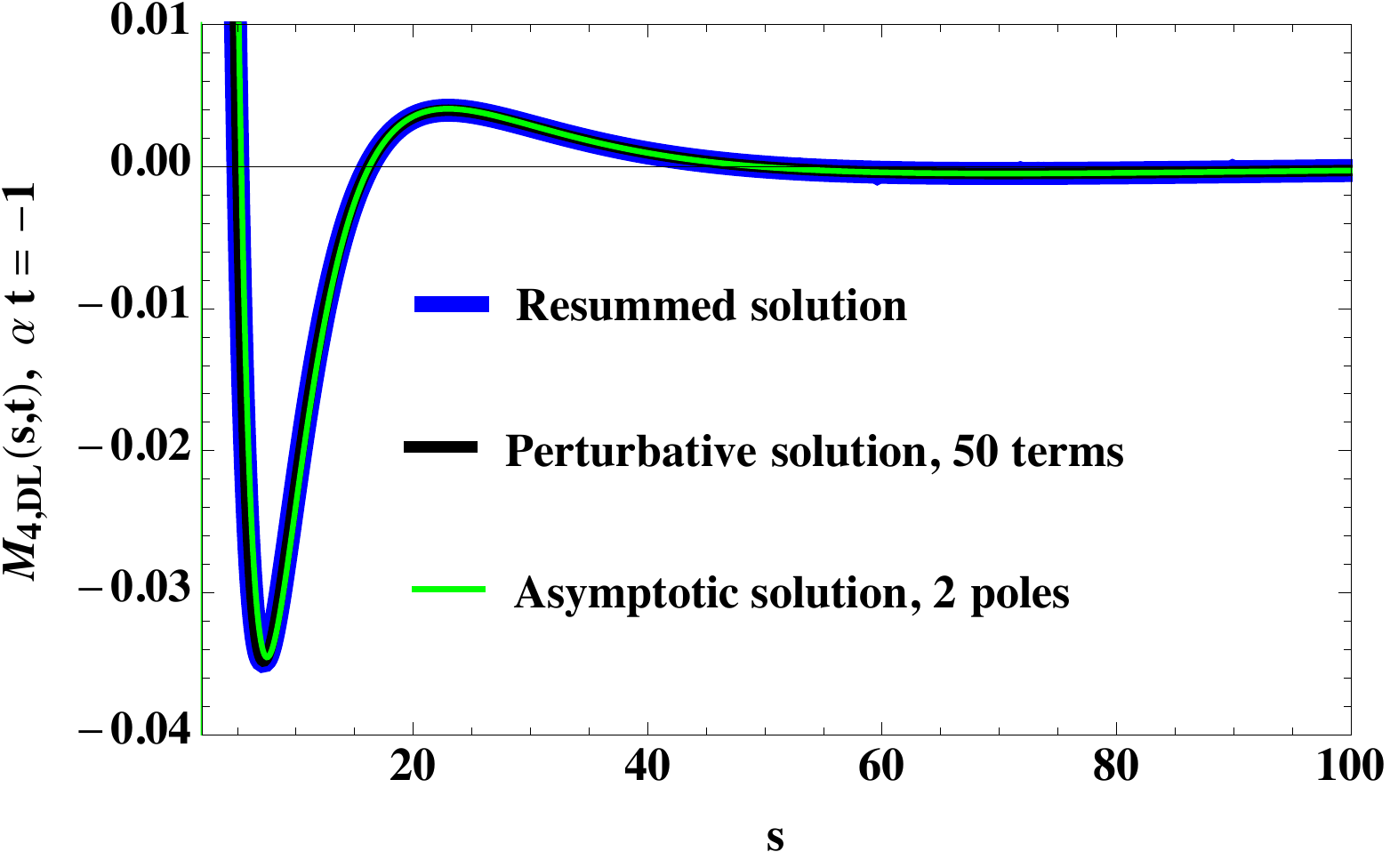}
\vspace{-.5cm}
\end{center}
\caption{The DL amplitude comparing the perturbative prediction with 50 terms to the full resummation in terms of the parabolic cylinder function. The approximation including two poles in the complex $\omega$-plane is also shown.}
\label{PertResumPlotb1}
\end{figure}

\begin{figure}
\hspace{-2cm}\includegraphics[width=19cm]{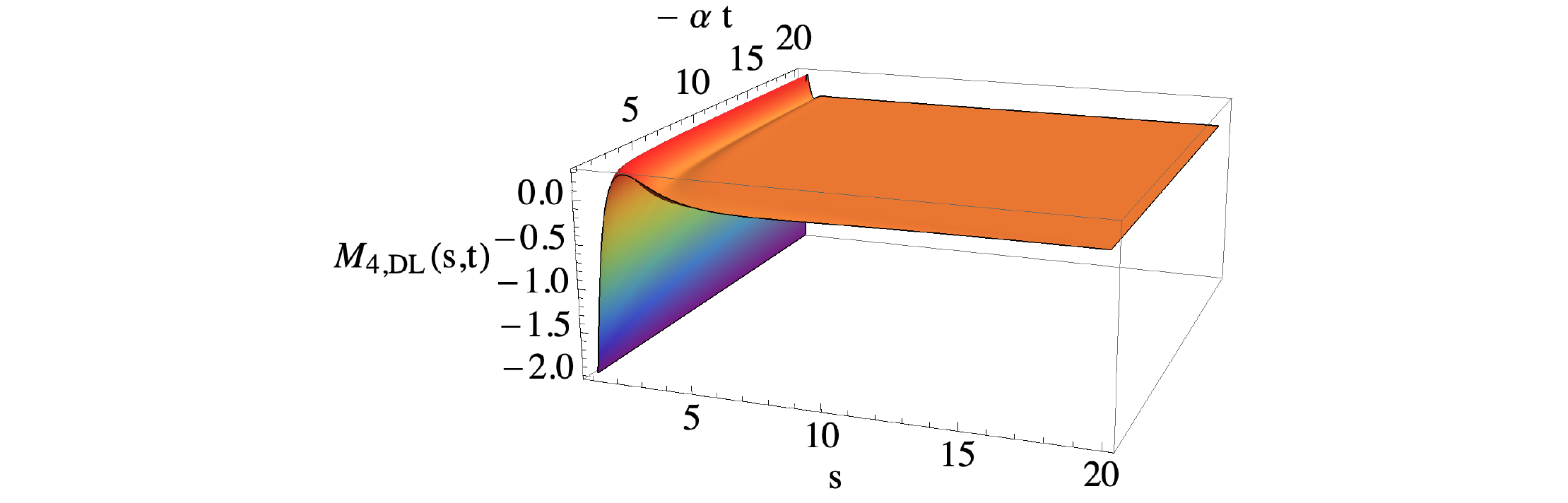}
\vspace{-.5cm}
\caption{The asymptotic DL amplitude as a function of $s$ and $-\alpha t$.}
\label{M4DLsversusat}
\end{figure}

A study of this result is included in Fig.~\ref{PertResumPlotb1}, where it can be seen that it numerically agrees with the perturbative expansion considering the first 50 terms in Eq.~(\ref{M4DLscamp}).  
 This expression is useful to understand the physical behaviour of the DL sector of the scattering amplitude as it will be shown in the next section. Both calculations agree with the numerical integration of the expression in 
Eq.~(\ref{omegaintegAmpl}) along the contour placed to the right of all the singularities in the integrand, which is also shown in Fig.~\ref{PertResumPlotb1}. The combined $s$ and $t$ dependence of Eq.~(\ref{AsympAmplitude}) is illustrated in Fig.~\ref{M4DLsversusat}.


\section{Impact parameter description}

\begin{figure}[h]
\begin{center}
\includegraphics[width=9.cm]{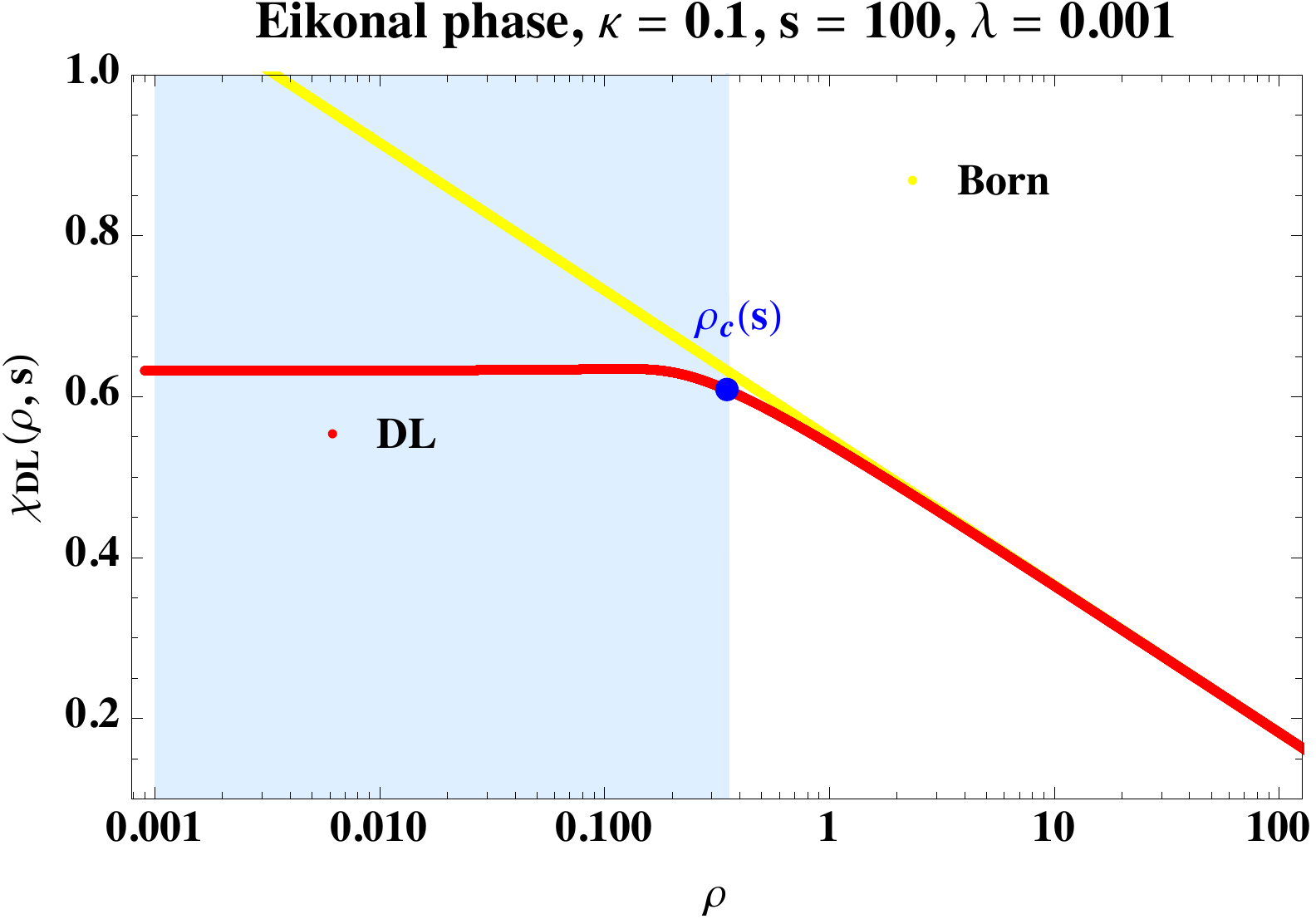}
\end{center}
\vspace{-.45cm}
\caption{Eikonal phase in Born and DL approximations as a function of 
the impact parameter $\rho$ with $\kappa = 0.1$, $s = 100$~GeV$^2$ and $\lambda = 0.001$ GeV.}
\label{deltaDLrho}
\end{figure}

To better understand the physical implications of the DL terms in the 4-graviton scattering amplitude it is helpful to work in the impact parameter representation. 
For this it is needed to make use of the following Fourier transform
\begin{eqnarray}
\frac{ \alpha  s^2 }{q^2} {\cal M}_{4,{\rm DL}}  (s,t) &=& \frac{  s }{4 \pi^2}
\int d^2 \vec{\rho} \, e^{i \vec{q} \cdot \vec{\rho}}  \chi_{\rm DL} \left(\rho,s\right) \nonumber\\
&\simeq& 
- \frac{ i s }{4 \pi^2}
\int d^2 \vec{\rho} \, e^{i \vec{q} \cdot \vec{\rho}} \left(e^{i \chi_{\rm DL} \left(\rho,s\right)}-1\right)
\label{eikdef}
\end{eqnarray}
where $\rho$ is the impact parameter for the graviton-graviton interaction, $q= \sqrt{-t}$, and an eikonal exponentiation has been introduced. The purpose of this section is to study the properties of the eikonal phase 
\begin{eqnarray}
\chi_{\rm DL} \left(\rho,s\right) &=&  \frac{\alpha s}{2} 
\int \frac{d^2 \vec{q}}{q^2} e^{- i \vec{q} \cdot \vec{\rho}} 
{\cal M}_{4,{\rm DL}}  (s,t) \, \theta\left(q - \lambda \right)  
\label{phaseikonal}
\end{eqnarray}
with DL accuracy  in the high energy limit. The first term in ${\cal M}_{4,{\rm DL}} $ is one and generates the Born one-graviton exchange contribution. Since 
$\alpha = \kappa^2 /(8 \pi^2) $ the Born eikonal phase reads
\begin{eqnarray}
\chi_{\rm Born} (\rho,s) &=& \frac{s \kappa^2}{8 \pi^2}
\int \frac{d^2 \vec{q}}{q^2} e^{-i \vec{q} \cdot \vec{\rho}} \, \theta\left(q - \lambda \right) ~ = ~ - \frac{s \kappa^2}{4 \pi} \ln{\left(\rho \, \lambda\right)}, 
\end{eqnarray}
where $\lambda$ is the IR regulator. The remaining terms in ${\cal M}_{4,{\rm DL}} (s,t)$ do not need of this regularization. To see this it is convenient to work with the asymptotic expression in Eq.~(\ref{AsympAmplitude}) keeping the contributions from the two complex-conjugated poles with the largest real part. The corresponding eikonal phase stemming from these poles, with the notation for their position being 
\begin{eqnarray}
\beta &=& \beta_1 \pm i \, \beta_2 ~=~ |\beta| \, e^{\pm i \beta_{\rm arg}},\\
\beta_1 &=& \frac{1.9159908576164295}{2 \sqrt{2} \pi} ~=~ 0.21562472883998976,\\
\beta_2 &=& \frac{2.8163594181520013}{2 \sqrt{2} \pi} ~=~ 0.31695179204062257,
\end{eqnarray}
is
\begin{eqnarray}
\lim_{s \to \infty} \chi_{\rm DL} \left(\rho,s\right) 
&=& \frac{s \kappa^2}{8 \pi^2} 
\int \frac{d^2 \vec{q}}{q^2} e^{- i \vec{q} \cdot \vec{\rho}} s^{- \beta_1 \kappa \, q}
\cos{ \left( \beta_2 \kappa \, q  \ln{s}\right)} \, \theta\left(q - \lambda \right) \nonumber\\
&=& -  \frac{s \kappa^2}{4 \pi} \ln\left(\frac{e^{\gamma} }{2}\lambda \, |\beta| \, \kappa \ln{s} 
 \left|1+\sqrt{1+\frac{\rho^2}{\beta^2 \kappa^2 \ln^2{s}}}\right| \right),
\end{eqnarray}
where $\gamma$ is the Euler's constant. From now on the convenient rescaling $\frac{e^\gamma \lambda}{2} \to \lambda$ is used. 
This expression has two distinct asymptotic regimes at small and large impact parameters,
\begin{eqnarray}
\lim_{s \to \infty, \rho \ll 1} \chi_{\rm DL} \left(\rho,s\right) &=&  -  \frac{s \kappa^2}{4 \pi} \ln{\left(2 \lambda \, |\beta| \, \kappa \ln{s}\right)},\\
\lim_{s \to \infty, \rho \gg 1} \chi_{\rm DL} \left(\rho,s\right) &=&   - \frac{s \kappa^2}{4 \pi} \ln{\left(\rho \, \lambda\right)}~=~ \chi_{\rm Born} (\rho,s),
\end{eqnarray}
which can be easily identified in Fig.~\ref{deltaDLrho}. 
The value of the impact parameter at which a strong departure from the Born behaviour  takes place can be found equating both expressions and corresponds to the energy dependent critical line $\rho_c (s) = 2 \, |\beta| \, \kappa \ln{s}$. Note that $\rho_c (s)$ is independent of the IR cutoff $\lambda$. 

\begin{figure}[h]
\begin{center}
\includegraphics[width=9cm]{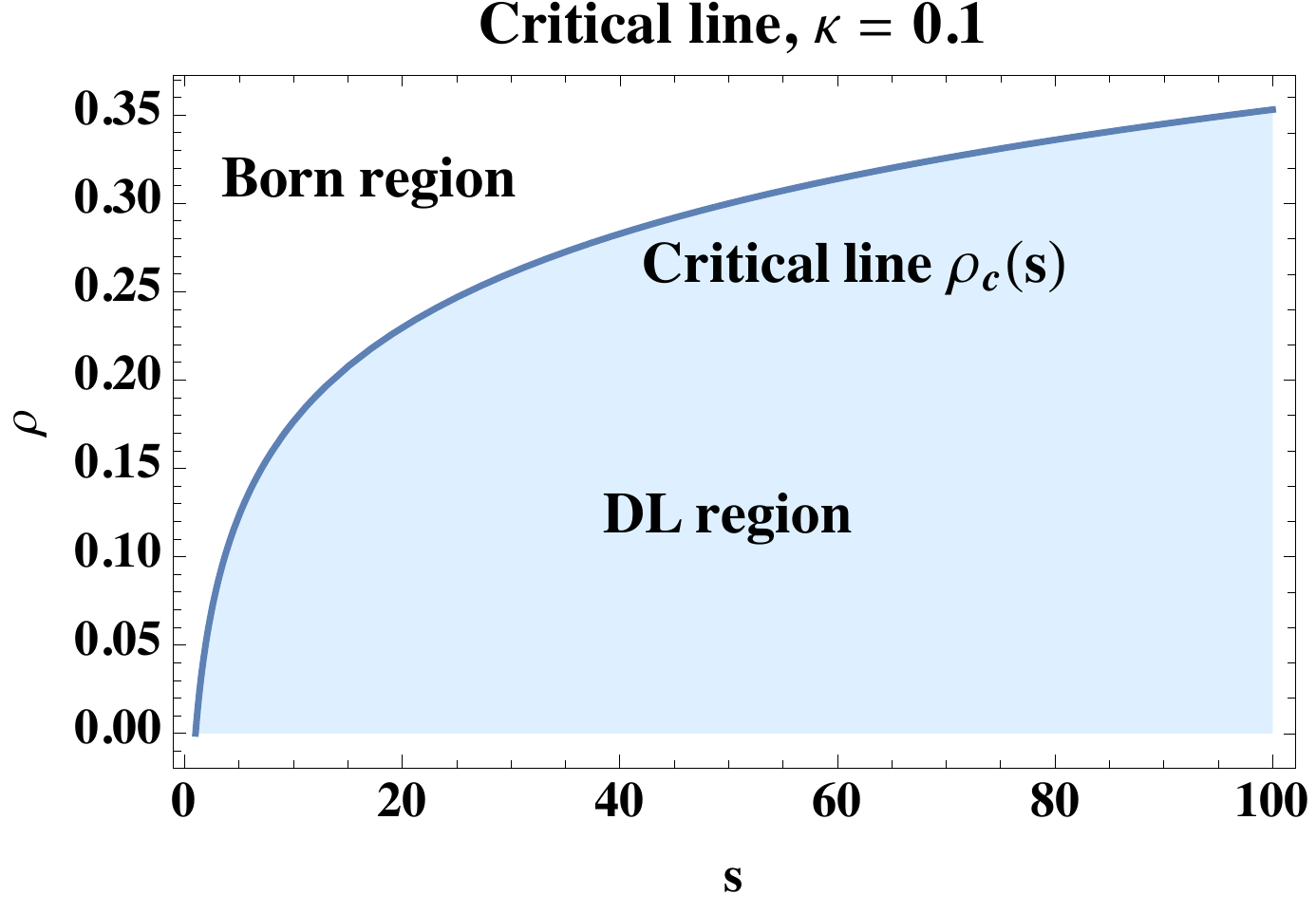}
\end{center}
\vspace{-.45cm}
\caption{Critical line in $\rho$ as a function of the center of mass energy. When $\rho < \rho_c (s)$ the double logs are important.}
\label{CriticalLine}
\end{figure}

In Fig.~\ref{deltaDLrho} there are two distinct regions well separated by a sharp transition: one dominated by the Born amplitude, at large $\rho >\rho_c (s)$, and a second one where the DL terms prevail, in the impact parameter region  $\rho < \rho_c (s)$.  In Fig.~\ref{CriticalLine} the logarithmic dependence with energy of the critical line for $\kappa=0.1$ is shown. In the region with impact factors  
below the critical line $\rho_c (s)$ the effect of the DLs is most important. Above this line the Born approximation dominates. 

\begin{figure}
\begin{center}
\includegraphics[width=8cm]{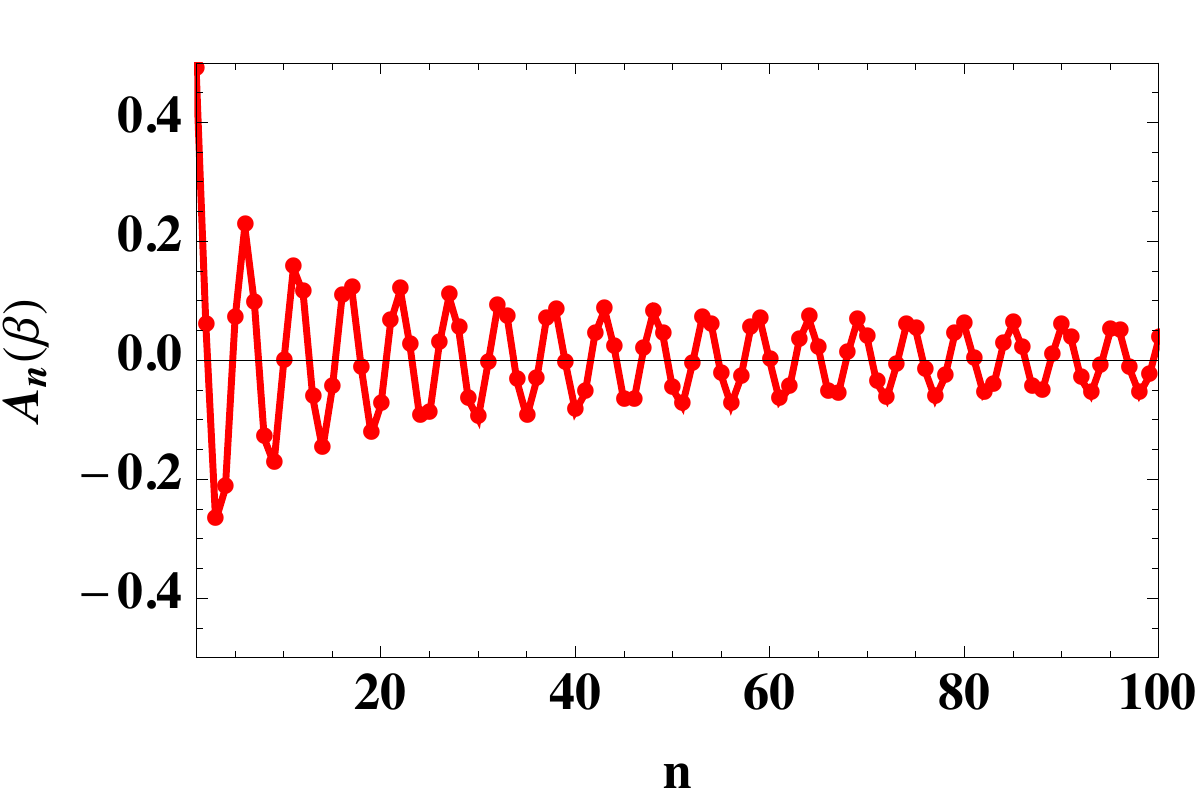}
\end{center}
\vspace{-.4cm}
\caption{Coefficients of asymptotic expansion ${\cal A}_n (\beta)$ in Eq.~(\ref{AnCoeffs}) as a function of $n$.}
\label{AnCoeffsplot}
\end{figure}

For small values of 
$\rho$ the natural expansion variable in the expression for the phase is 
$2 \rho/\rho_c(s)$. More precisely, 
\begin{eqnarray}
\lim_{s \to \infty, 2 \rho< \rho_c} \chi_{\rm DL} \left(\rho,s\right)&=& -  \frac{s \kappa^2}{4 \pi} \ln\left(\lambda \frac{\rho_c(s)}{2} 
 \left|1+\sqrt{1+\frac{\beta}{\beta^*}\frac{4\rho^2 }{\rho_c^2(s)}}\right|\right)
 \nonumber\\ 
 &\simeq& 
-  \frac{s \kappa^2}{4 \pi} \ln{\left(\lambda \rho_c (s)\right)} 
+ \frac{s \kappa^2}{4 \pi}  \sum_{n=1}^\infty  \frac{{\cal A}_n (\beta)}{2 n}  \left(\frac{2   \rho}{ \rho_c (s)}\right)^{2n},
\end{eqnarray}
where 
\begin{eqnarray}
{\cal A}_n (\beta) &=& \frac{(-1)^n }{ \sqrt{\pi} n!} 
\Gamma \left(\frac{1}{2}+n\right)
T_{2 n}(\cos (\beta_{\rm arg} )).
\label{AnCoeffs}
\end{eqnarray}
$\Gamma (z)$ is the Gamma function and $T_{n}(\cos (\theta)) = \cos{(n \theta)}$ are the Chebyshev polynomials of the first kind. This is an asymptotic expansion which works accurately when $2 \rho < \rho_c (s)$. The coefficients ${\cal A}_n (\beta)$ in Eq.~(\ref{AnCoeffs}) are shown as a function of $n$ in 
Fig.~\ref{AnCoeffsplot}.  

It is more common to use expansions at large impact parameter which show an order by order departure from the well known Born behaviour (see, {\it e.g.},~\cite{Collado:2018isu}). In the present case the asymptotic series for large impact parameter,  $2 \rho > \rho_c (s)$, is
\begin{eqnarray}
\lim_{s \to \infty, 2 \rho >\rho_c} \chi_{\rm DL} \left(\rho,s\right) 
&=&  \chi_{\rm Born} (\rho,s) 
-  \frac{s \kappa^2}{4 \pi} \sum_{n=0}^\infty \frac{{\cal A}_n (\beta)}{2n+1 } 
\left(\frac{\rho_c(s)}{2 \rho}\right)^{2n+1}.
\end{eqnarray}
\begin{figure}
\begin{center}
\includegraphics[width=11cm]{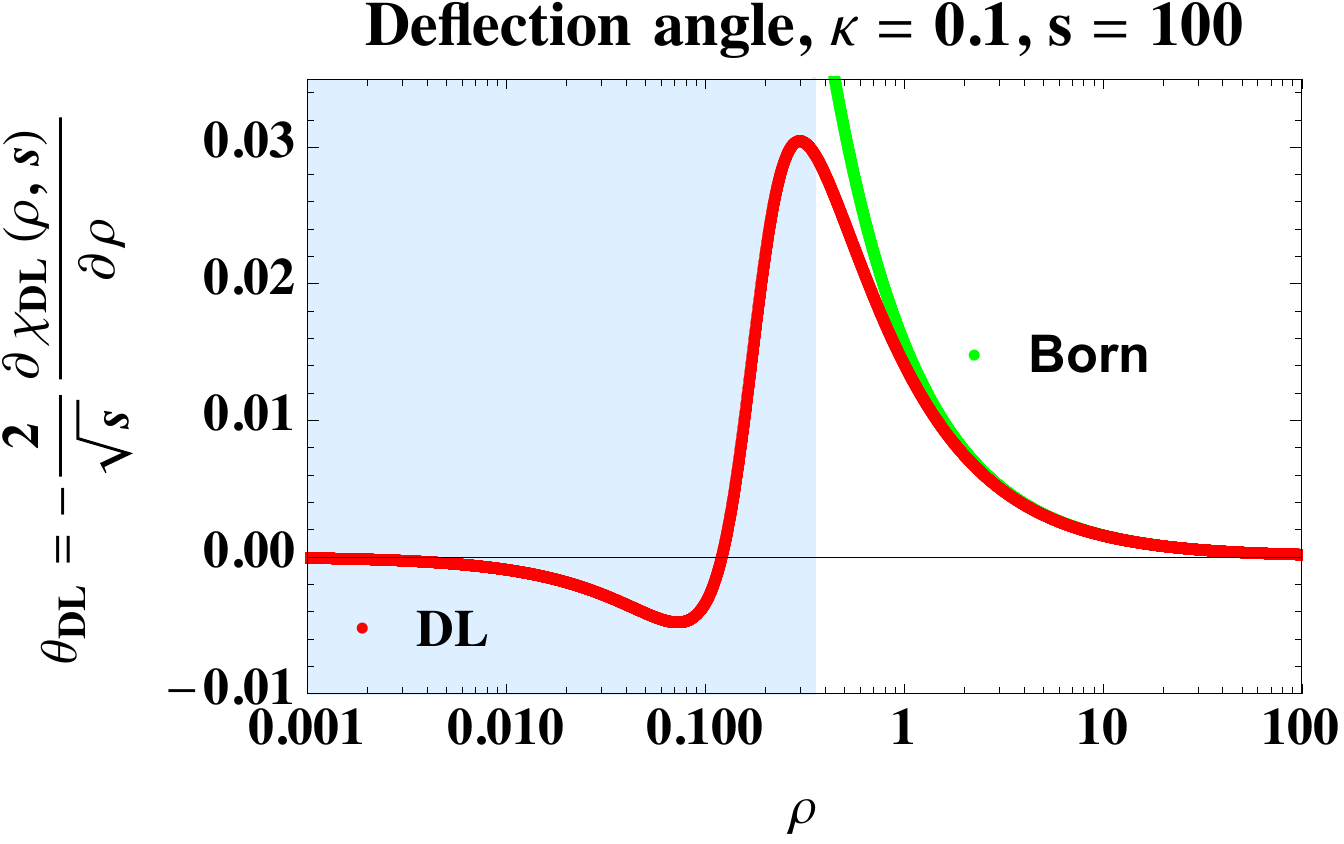}
\end{center}
\vspace{-.1cm}
\caption{Deflection angle in Born and DL approximations as a function of 
$\rho$ with $\kappa = 0.1$, $s = 100$~GeV$^2$}
\label{DeflectionAngle}
\end{figure}
\begin{figure}
\begin{center}
\includegraphics[width=11cm]{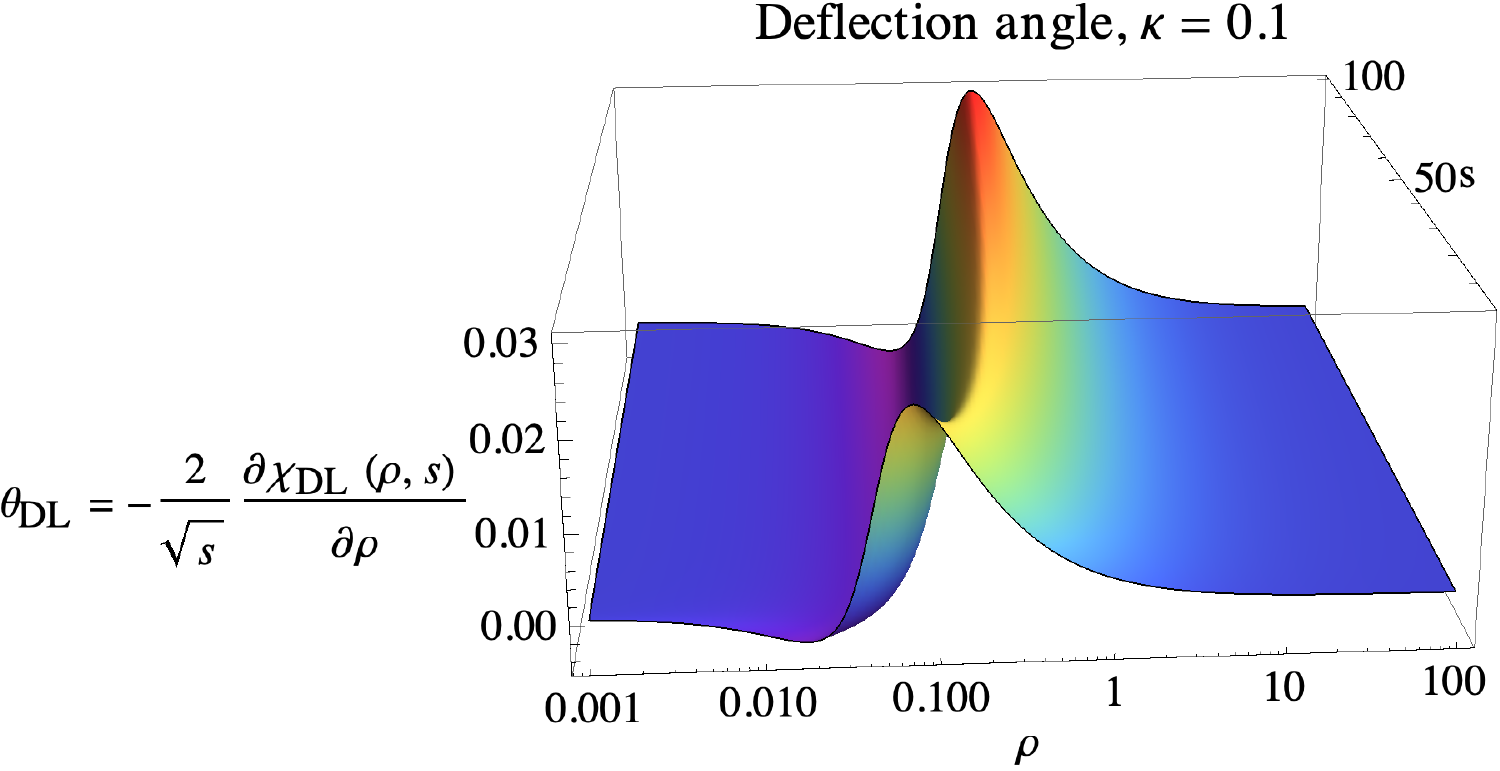}
\end{center}
\vspace{-.4cm}
\caption{Deflection angle in DL approximation as a function of 
$\rho$ and $s$.}
\label{DeflectionAngle3D}
\end{figure}

A well known test of Einstein's general theory of relativity is the bending of 
light when passing near an object with large mass. Recently, this has been studied for gravitons in the presence of a massive scalar particle and making use of one-loop amplitudes~\cite{Chi:2019owc}. In the present context it is possible to find DL quantum corrections to a similar bending in the case of graviton-graviton interaction. The small center-of-mass scattering angle limit, $\theta \ll 1$, with $s>0$ and $t = -\frac{s}{2} (1-\cos{\theta}) \leq0$, implies 
$t = - q^2  \simeq - s \frac{\theta^2}{4}$. Hence 
$q \simeq \sqrt{s} {\theta \over 2} \ll 1$ and, with the integration in Eq.~(\ref{eikdef}) dominated by the stationary phase defined by 
\begin{eqnarray}
\frac{\partial}{\partial \rho} \left(q \rho + \chi_{\rm DL} (s,\rho)\right)=0,
\end{eqnarray}  
the contribution from the DLs to the deflection angle corresponds  to
\begin{eqnarray}
\lim_{s \to \infty} \theta_{\rm DL} &=& -\frac{2}{\sqrt{s}}\frac{\partial \chi_{\rm DL} \left(\rho,s\right)}{\partial \rho} 
~ = ~  \frac{\sqrt{s} \, \kappa^2}{2 \pi \rho} \Re \left(1-\frac{1}{\sqrt{1+\frac{\beta}{\beta^*}\frac{4\rho^2 }{\rho_c^2(s)}}}  \right).
\end{eqnarray}

This angle is plotted in Fig.~\ref{DeflectionAngle} for $\kappa = 0.1$ and $s=100$. Its dependence with $s$ is also included in Fig.~\ref{DeflectionAngle3D}. It is worth noting the change of sign in the angle which can be interpreted as a modification of gravity at small impact parameters.  Up to $\rho = \rho_c (s)$ the attractive nature of gravity at large impact parameters is manifest. For distances smaller than $\rho_c(s)$ the gravitinos present in the DL quantum corrections at high energies generate a repulsive contribution to the graviton-graviton interaction. This screening correction to the gravitational potential is associated to the negative sign in the quadratic term of Eq.~(\ref{fomegqeq}). 

\begin{figure}
\begin{center}
\includegraphics[width=11cm]{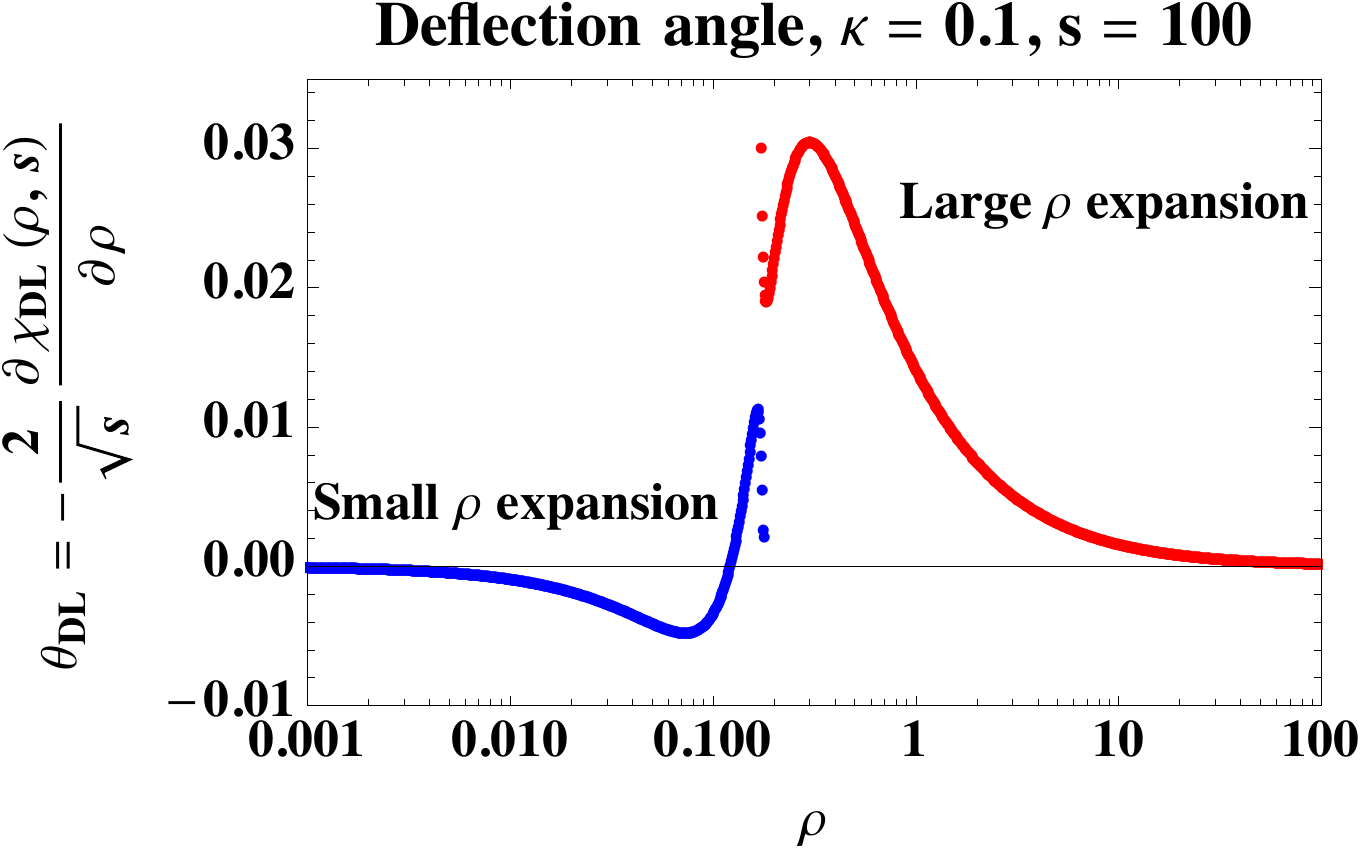}
\end{center}
\vspace{-.1cm}
\caption{Asymptotic expansions for the deflection angle in the DL approximation as a function of $\rho$ with $\kappa = 0.1$, $s = 100$~GeV$^2$.}
\label{SmallLargeRho}
\end{figure}

In terms of asymptotic expansions at large and small impact parameters the deflection angle can be written in the form
\begin{eqnarray}
\lim_{s \to \infty} \theta_{\rm DL}  
&=&   \frac{\sqrt{s} \, \kappa^2}{2 \pi  \rho}\Bigg[ \left(1-  \sum_{n=0}^\infty {\cal A}_n (\beta)
\left(\frac{\rho_c(s)}{2 \rho}\right)^{2n+1} \right)
\theta \left(\rho - \frac{\rho_c(s)}{2}\right) \nonumber\\
&&\hspace{2cm}-    \sum_{n=1}^\infty  {\cal A}_n (\beta)
  \left(\frac{2  \rho}{\rho_c(s)}\right)^{2n}\theta \left( \frac{\rho_c(s)}{2}-\rho \right) \Bigg]. 
\end{eqnarray}
This expansion gives an accurate description of the DL contributions to the graviton's semiclassical trajectory bending at high energies away from the region $\rho \simeq \rho_c (s)/2$ as it can be seen in Fig.~\ref{SmallLargeRho}.

\section{Relation to rooted ribbon graphs}

In this final section a one-to-one mapping between the evolution equation for 
the $t$-channel partial wave 
$f_\omega$ and the theory of graphs embedded on orientable surfaces is established. The starting point is the work of Arqu{\`e}s and B{\'e}raud in Discrete Mathematics where they investigated the so-called rooted maps on orientable surfaces~\cite{Arques:2000Ar}. 

A topological map on an compact orientable surface is a partition of the surface with a set of vertices, a set of edges connecting pairs of vertices or a vertex to itself, and a set of faces whose boundaries are the vertices and edges. These faces are homeomorphic to an open disc. The genus of the map corresponds to the genus of the surface on which it is embedded. The edges can be ``split" into two half-edges with a fixed orientation and there exist different ways they can enter a vertex. In the literature this splitting of an edge into two half-edges is denoted as the generation of a ``fat"  or ``ribbon" graph. 

In a ribbon graph the vertices are represented by small circles and edges by ribbons. The two half-edges of a ribbon are each labelled by permutations $\alpha, \sigma \in S_{2e}$, with $2 e$ being the number of half-edges. In order to avoid unnecessary redundancies one can remove automorphisms of the graph by making one half-edge distinct assigning to it an arrow, or root. The vertex incident to this root is called the root vertex. The connection to the results here presented  appears when solving the problem of enumerating equivalence classes of isomorphic rooted maps independently of their genus, {\it i.e.} only considering the number of vertices and edges. 

When calculating the generating series of these maps in~\cite{Arques:2000Ar} the same differential equation as the one written for the $t$-channel partial wave $f_\omega$ in  Eq.~(\ref{fomegqeq}) was obtained. In their notation $M(y,z)$ corresponds to the generating series of orientable rooted maps of any genus with the exponent of $y$ being the number of vertices and the exponent 
of $z$ the number of edges in the map and follows the topological equation
 \begin{eqnarray}
 M (y,z) &=& y + z \left(M^2 (y,z) +  M(y,z) + 2 z \frac{\partial}{\partial z}M(y,z) \right).
 \end{eqnarray}
 As a corollary, $y=1$ sets the equation for the generating series of 1-rooted maps with respect to the number of edges 
 \begin{eqnarray}
 M (z) &=& 1 + z \left(M^2 (z) +  M(z) + 2 z \frac{d}{d z}M(z)\right),
 \end{eqnarray}
 where the notation $M(1,z) \equiv M(z)$  has been used. This is exactly Eq.~(\ref{fomegqeq}) with the relations
 \begin{eqnarray}
 f_\omega = M(z) = \sum_{n=0}^\infty {\cal C}_n z^n, \, \, \, z = \frac{\alpha t}{\omega^2}, 
  \end{eqnarray}
In terms of the scattering amplitude $2n$ is the exponent of the logarithms of energy while in the generating function for the enumeration of ribbon graphs it indicates the number of half-edges. In gravity the ${\cal C}_n$ coefficients are obtained after summing all the possible contributions from different Feynman diagrams. The alternative interpretation is that ${\cal C}_n$ is the number of single rooted graphs with $n$ edges regardless of genus. 
 
A novel graphical bijection between the Feynman and ribbon diagrams can be drawn. A comprehensive review on connections to other types of Feynman diagrams can be found in~\cite{Prunotto:2013tpa} and recent related works 
in~\cite{Gopala:2017mjh,GopalaKrishna:2018wzq}. The tree level diagram corresponds to the one root ribbon graph with zero edges which is a degenerate case. This diagram is considered to be a rooted one. Graphically,
  \begin{eqnarray}
 \hspace{-.cm}\parbox{15mm}{\includegraphics[width=2.2cm,angle=0]{Fw-tree}} \hspace{.8cm}
 &=& \hspace{.3cm} \parbox{15mm}{\includegraphics[width=.6cm,angle=0]{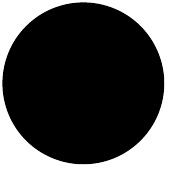}} 
\end{eqnarray}
In SUGRA, the DLs are generated by attaching graviton and gravitino propagators to this basic graph. With one edge, the one root ribbon graphs combine into two distinct forms. The first one, for a single vertex, sets the correspondence
\begin{eqnarray}
   2 \hspace{.cm} \parbox{15mm}{\includegraphics[width=2.2cm,angle=0]{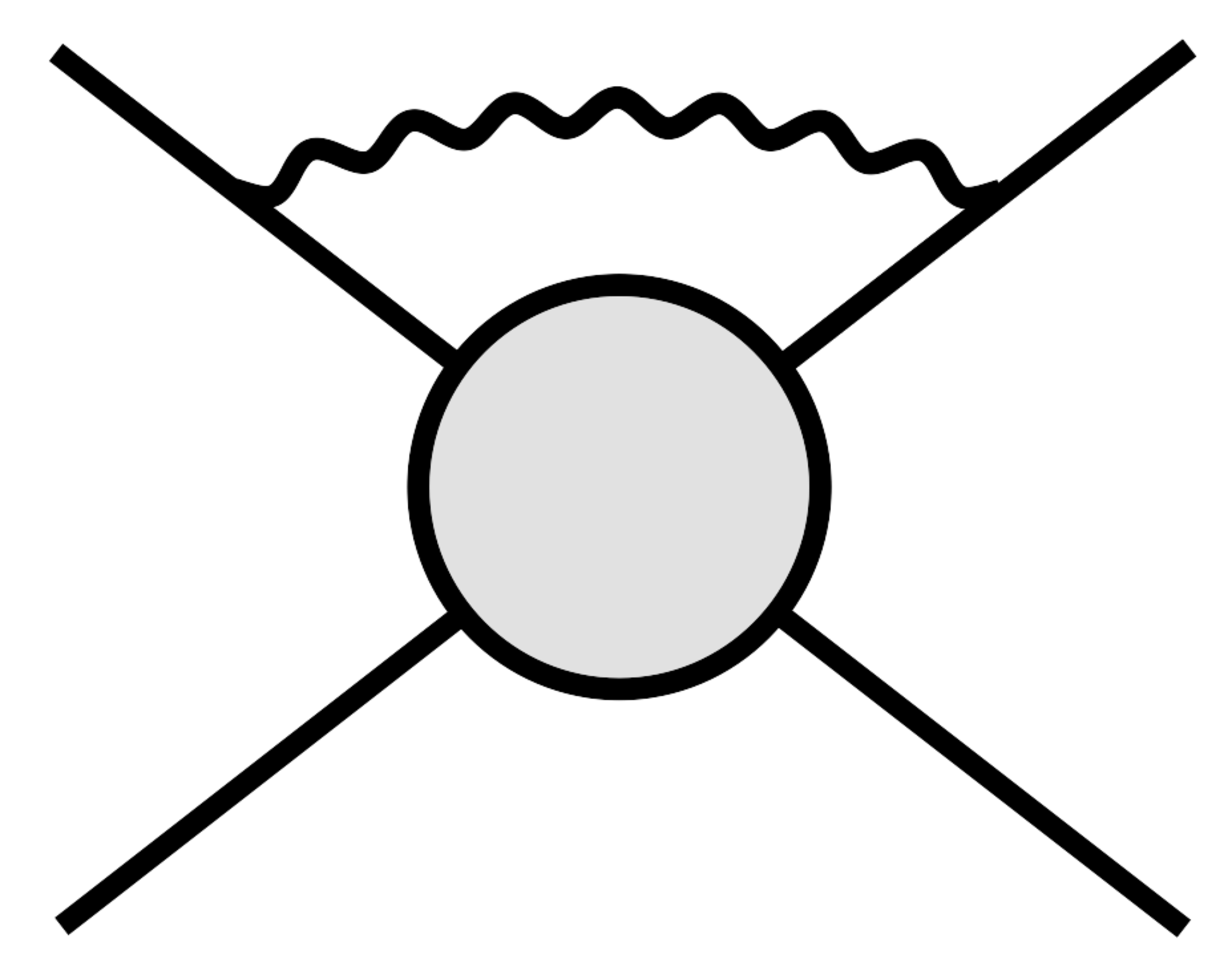}}
 \hspace{0.5cm}
 + 2 \hspace{.cm} \parbox{15mm}{\includegraphics[width=2.2cm,angle=0]{Fw-soft-right}}
 \hspace{0.8cm}
  &=& \hspace{.3cm} \parbox{15mm}{\includegraphics[width=1.4cm,angle=0]{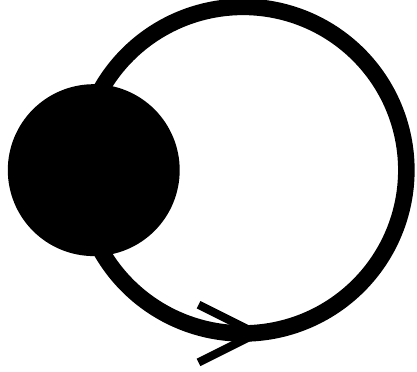}}\hspace{1cm}\hspace{-.8cm}
\end{eqnarray}
In a single ribbon graph the two field theory topologies, planar and non-planar, are included. The second one maps the exchange in the $t$-channel of two gravitons and two gravitinos with a two-vertex 1-rooted ribbon map:
  \begin{eqnarray}
 \hspace{-.cm}\parbox{15mm}{\includegraphics[width=2.8cm,angle=0]{Fw-Fw}} \hspace{1.4cm}
 &=& \hspace{.3cm} \parbox{15mm}{\includegraphics[width=2.8cm,angle=0]{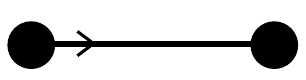}} 
\end{eqnarray}
These two possible configurations define the second term in the series of coefficients $(1,2,10,74,\dots)$ in Eq.~(\ref{fullpertexp}). Iterating the original equation all the coefficients are generated order by order in the number of edges. As a further example, the 10 graphs with two  edges correspond to all the possible combinations of the previous two ones, {\it i.e.}
  \begin{eqnarray}
 \hspace{-3.cm}\parbox{30mm}{\includegraphics[width=4.2cm,angle=0]{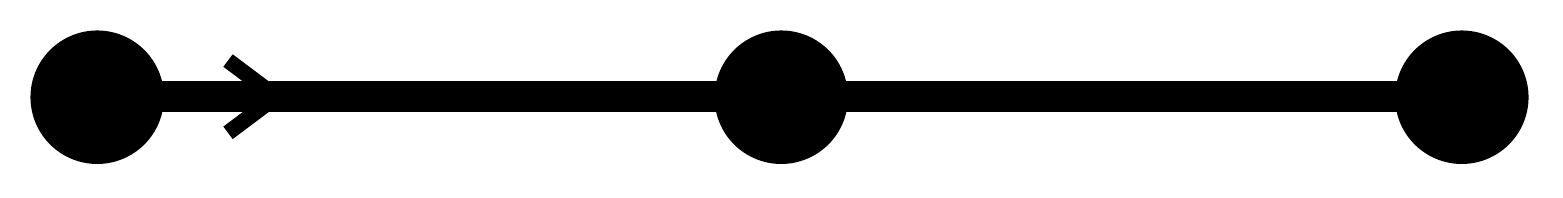}} \hspace{1.4cm}
 && \hspace{.3cm} \parbox{30mm}{\includegraphics[width=4.2cm,angle=0]{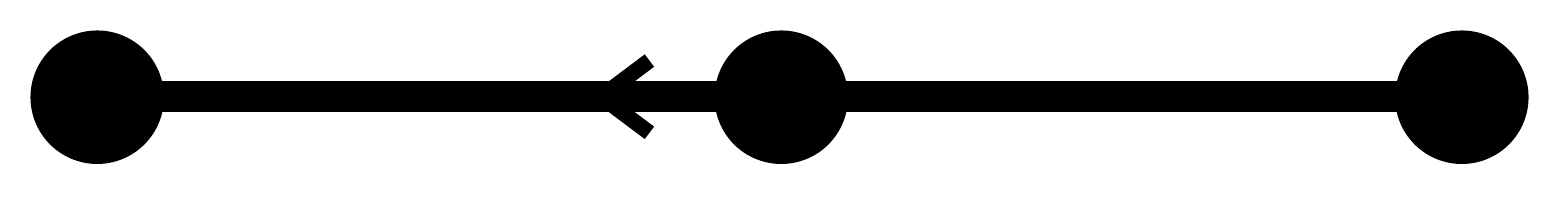}} \nonumber\\
 \hspace{.4cm}\parbox{25mm}{\includegraphics[width=3.1cm,angle=0]{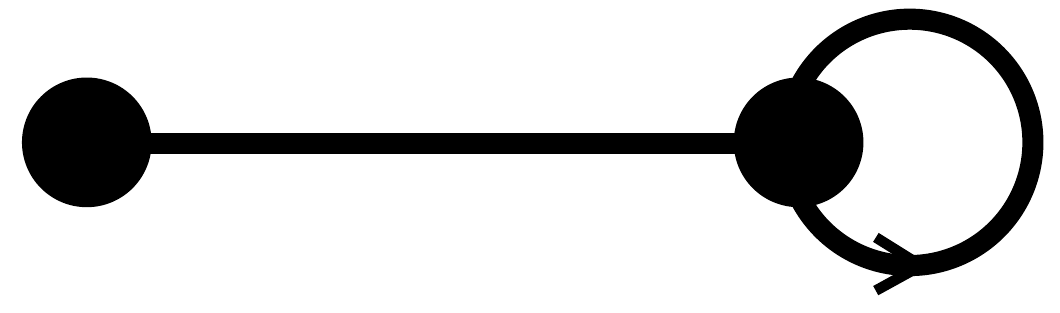}} \hspace{1.4cm}
 && \hspace{.8cm} \parbox{25mm}{\includegraphics[width=3.1cm,angle=0]{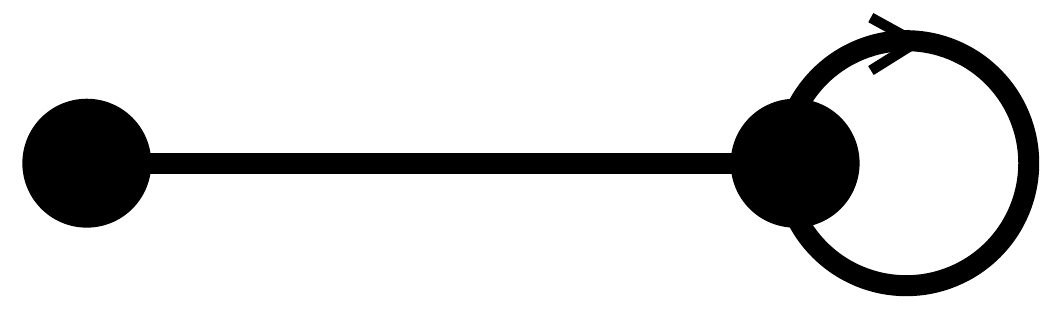}} \nonumber\\
  \hspace{-.cm}\parbox{25mm}{\includegraphics[width=3.1cm,angle=0]{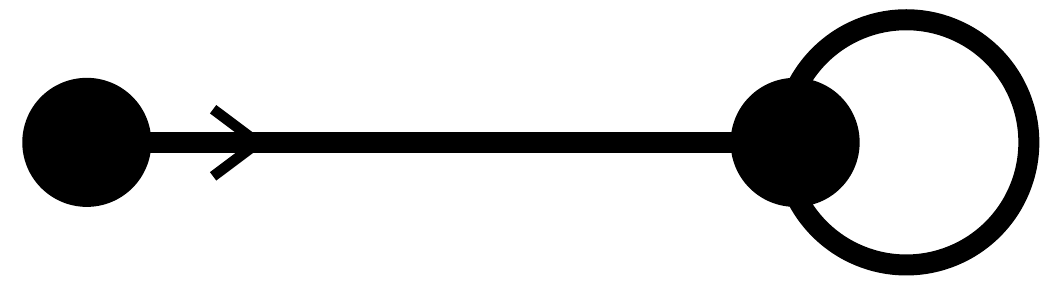}} \hspace{1.4cm}
 && \hspace{.8cm} \parbox{25mm}{\includegraphics[width=3.1cm,angle=0]{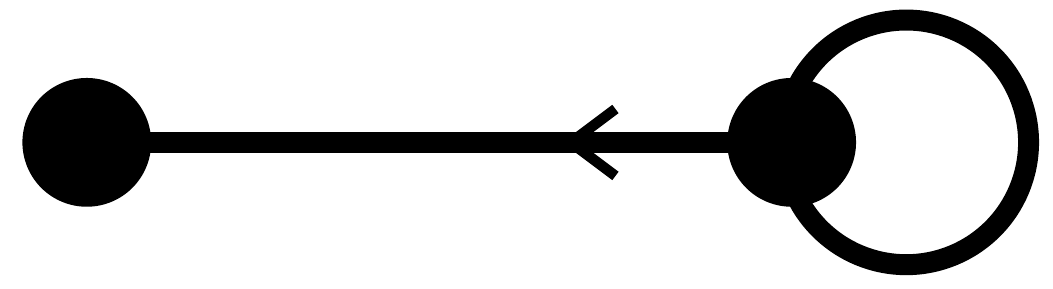}} \nonumber\\
  \hspace{-.cm}\parbox{19mm}{\includegraphics[width=1.8cm,angle=0]{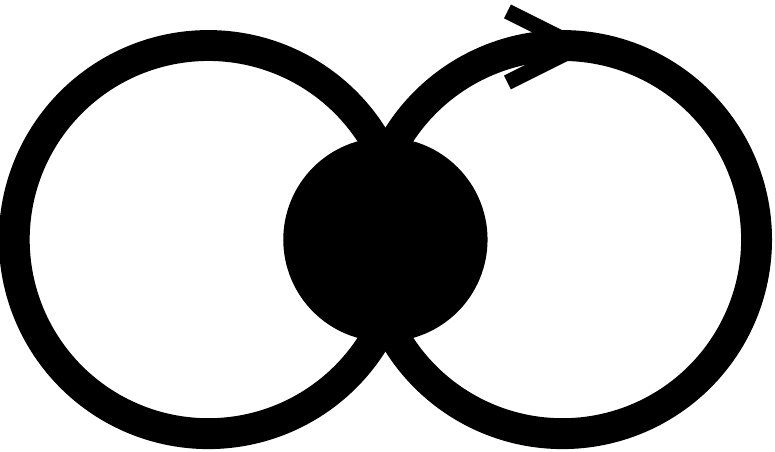}} \hspace{1.4cm}
 && \hspace{1.4cm} \parbox{19mm}{\includegraphics[width=1.8cm,angle=0]{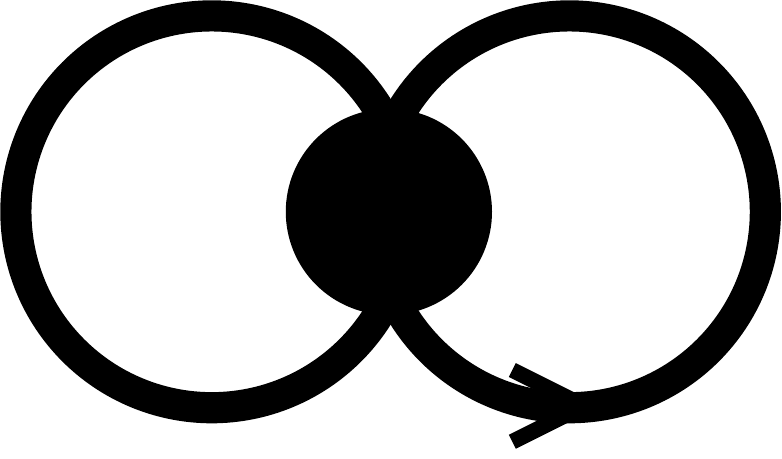}} \nonumber\\
  \hspace{-.2cm}\parbox{16.mm}{\includegraphics[width=1.3cm,angle=0]{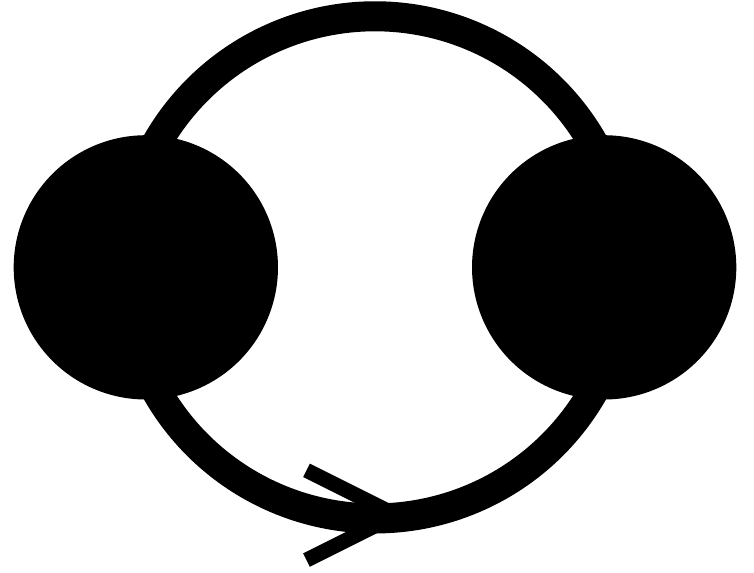}} \hspace{1.4cm}
 && \hspace{1.5cm} \parbox{15mm}{\includegraphics[width=1.6cm,angle=0]{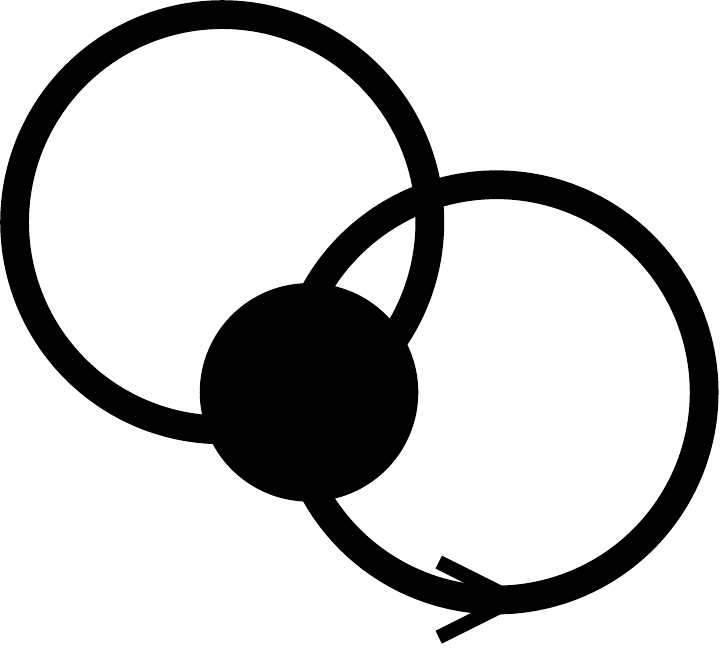}} \nonumber
\end{eqnarray}
This representation is more economic than the Feynman diagram approach.  At each order of perturbation theory the multiplicity of possible 1-rotted ribbon graphs and the corresponding DL coefficients of the ${\cal N}$$=$ 8 SUGRA 4-graviton scattering amplitude grow very fast. As previously discussed, they admit the representation
\begin{eqnarray}
{\cal C}_n &=&  {2^{n+1}  \over \pi^{\frac{3}{2}}}   
\int_0^\infty \frac{x^{n-\frac{1}{2}} e^{x}}{1+{\rm erfi}^2 (\sqrt{x})}dx.
\label{CCn}
\end{eqnarray}

\begin{figure}
\begin{center}
\includegraphics[width=7cm]{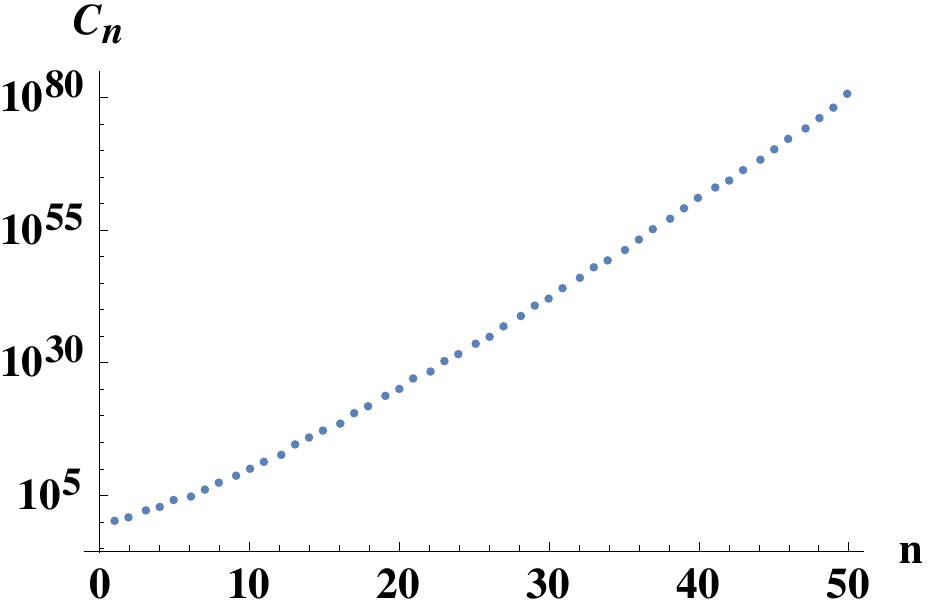}\\
\vspace{.5cm}
\includegraphics[width=7cm]{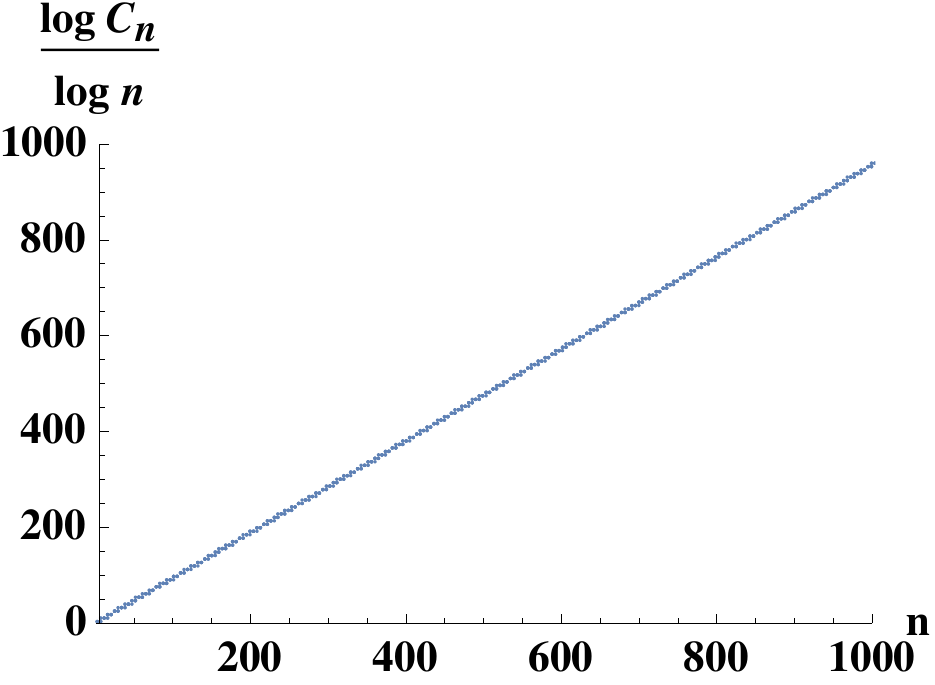}
\end{center}
\vspace{-.1cm}
\caption{Coefficients ${\cal C}_n$, which correspond to the number of 1-rooted diagrams with $n$ edges, as a function of $n$. }
\label{Cn}
\end{figure}
Their growth with $n$ can be seen in Fig.~\ref{Cn}. In the lower plot of that figure the approximate linear growth with $n$ of the function $\ln{\cal C}_n /\ln{n}$ emerges. At large $n$ the growth for the number of 1-rooted diagrams behaves as ${\cal C}_n \simeq n^n$. A more precise analysis for $n\to \infty$ gives
\begin{eqnarray}
\lim_{n \to \infty} {\cal C}_n 
&\simeq&  {2^{n+1}  \over \sqrt{\pi}}   
\int_0^\infty  x^{n+\frac{1}{2}} e^{-x}dx
~=~  {2^{n+1}  \over \sqrt{\pi}}   \Gamma \left(n+\frac{3}{2}\right) 
~\simeq~ \sqrt{2} \, e^{-n} (2n)^{n+1}.
\end{eqnarray}
The integrand in Eq.~(\ref{CCn}) has a sharp maximum at $x\simeq n$. Therefore, when $n$ is large it can be approximated by the $x \gg1$ expansion  $e^{2x}/(1+{\rm erfi}^2 (\sqrt{x})) \simeq \pi x$ to obtain this asymptotic result. 

\section{Conclusions}

A new representation for the double-logarithmic (DL) terms $(\alpha \, t \ln^2 {s})^n$ in the 4-graviton scattering amplitude in ${\cal N}$$=$ 8 supergravity (SUGRA) to all orders has been presented in Eq.~(\ref{M4DLscamp}). This improves the results in~\cite{Bartels:2012ra}. The terms with $n \leq 3$ are in agreement with the complete amplitude calculations and the higher order ones serve as a non-trivial check for future exact results.  

Beyond a purely perturbative approach, the singularity structure in the complex angular momentum $\omega$-plane of the $t$-channel partial wave associated to the DL contributions to the amplitude, $f_\omega$ in Eq.~(\ref{M4Mellin}), has been investigated in detail. Its singularities are an infinite set of simple poles positioned asymptotically close to two lines at the left hand side of the $\omega$-plane, see Fig.~\ref{ParabCylZeroesList}. The DL asymptotics is therefore dominated by the two poles with the largest real part and allows for a simple representation of the amplitude which is given in Eq.~(\ref{AsympAmplitude}). It is numerically equivalent to the perturbative and exact results at high energies, see Fig.~\ref{PertResumPlotb1}. 

Making use of the asymptotic expression calculated from the leading pole singularities, the 4-graviton amplitude has been studied in impact parameter,  $\rho$, representation. A critical line $\rho = \rho_c (s) \simeq \sqrt{\alpha} \ln{s}$
appears such that when the impact parameter in the graviton-graviton interaction is below $\rho_c$ the DL contributions generate a repulsive contribution to the gravitational potential. Such screening of the gravitational interaction can be traced back to the gravitino sector of the theory and is associated to the 
convergent behaviour of the DL limit in the $2 \to 2$ amplitude as $s$ increases. In~\cite{Bartels:2012ra} it was shown that the DL sector of the 4-graviton amplitude is less convergent when moving from ${\cal N}$$=$ 8 to lower values of ${\cal N}$. Work is in progress to find out how this translates into the $\rho$ picture here presented.  

A mapping between the partial wave for the DL contributions to the 4-graviton scattering amplitude in ${\cal N}$$=$ 8 SUGRA and the generating function for the number of 1-rooted ribbon maps on orientable surfaces has been found. It will be interesting to deepen the understanding of this correspondence and extend it to other SUGRAs.

\section*{Acknowledgements} 

Research supported by the Spanish Research Agency (Agencia Estatal de Investigaci{\' o}n) Grant FPA2016-78022-P and IFT Centro de Excelencia Severo Ochoa Grant SEV-2016-0597.

\end{document}